\documentclass[sigconf]{acmart}

\usepackage{tcolorbox}
\usepackage{algorithm}
\usepackage{algorithmic}

\usepackage{tabularx}

\usepackage{amsmath}
\usepackage{amssymb}
\usepackage{graphicx}
\usepackage{booktabs}  
\usepackage{threeparttable} \usepackage{multirow} 
\usepackage{color}

\usepackage{subfigure}

\AtBeginDocument{%
  }

\copyrightyear{2025}
\acmYear{2025}
\setcopyright{acmlicensed}\acmConference[SIGIR '25]{Proceedings of the 48th International ACM SIGIR Conference on Research and Development in Information Retrieval}{July 13--18, 2025}{Padua, Italy}
\acmBooktitle{Proceedings of the 48th International ACM SIGIR Conference on Research and Development in Information Retrieval (SIGIR '25), July 13--18, 2025, Padua, Italy}
\acmDOI{10.1145/3726302.3730058}
\acmISBN{979-8-4007-1592-1/2025/07}

\begin{document}

\title{PR-Attack: Coordinated Prompt-RAG Attacks on Retrieval-Augmented Generation in Large Language Models via Bilevel Optimization}

\author{Yang Jiao}
\affiliation{%
  \institution{Tongji University}
  \city{Shanghai}
  \country{China}  
}
\email{yangjiao@tongji.edu.cn}

\author{Xiaodong Wang}
\affiliation{%
  \institution{Columbia University}
  \city{New York}
  \country{USA}}
\email{wangx@ee.columbia.edu}

\author{Kai Yang}
\authornote{Corresponding author.}
\affiliation{%
  \institution{Tongji University}
  \city{Shanghai}
  \country{China}
}
\email{kaiyang@tongji.edu.cn}

\renewcommand{\shortauthors}{Yang Jiao, Xiaodong Wang, and Kai Yang}

\begin{abstract}
 Large Language Models (LLMs) have demonstrated remarkable performance across a wide range of applications, e.g., medical question-answering, mathematical sciences, and code generation. However, they also exhibit inherent limitations, such as outdated knowledge and susceptibility to hallucinations. Retrieval-Augmented Generation (RAG) has emerged as a promising paradigm to address these issues, but it also introduces new vulnerabilities. Recent efforts have focused on the security of RAG-based LLMs, yet existing attack methods face three critical challenges: (1) their effectiveness declines sharply when only a limited number of poisoned texts can be injected into the knowledge database, (2) they lack sufficient stealth, as the attacks are often detectable by anomaly detection systems, which compromises their effectiveness, and (3) they rely on heuristic approaches to generate poisoned texts, lacking formal optimization frameworks and theoretic guarantees, which limits their effectiveness and applicability. To address these issues, we propose  coordinated Prompt-RAG attack (PR-attack), a novel optimization-driven attack that introduces a small number of poisoned texts into the knowledge database while embedding a backdoor trigger within the prompt. When activated, the trigger causes the LLM to generate pre-designed responses to targeted queries, while maintaining normal behavior in other contexts. This ensures both high effectiveness and stealth. We formulate the attack generation process as a bilevel optimization problem leveraging a principled optimization framework to develop optimal poisoned texts and triggers. Extensive experiments across diverse LLMs and datasets demonstrate the effectiveness of PR-Attack, achieving a high attack success rate even with a limited number of poisoned texts and significantly improved stealth compared to existing methods. These results highlight the potential risks posed by PR-Attack and emphasize the importance of securing RAG-based LLMs against such threats.
\end{abstract}

\ccsdesc[500]{Information systems~Adversarial retrieval}

\keywords{Retrieval-Augmented Generation, Large Language Models, Bilevel Optimization}


\maketitle

\section{Introduction}
Large Language Models (LLMs) have exhibited exceptional performance across a broad spectrum of applications, such as medical question-answering \cite{maharjan2024openmedlm}, chemical research \cite{boiko2023autonomous}, and mathematical sciences \cite{romera2024mathematical}. Prompt learning plays a crucial role in enhancing the adaptability of LLMs to various downstream tasks \cite{du2022ppt,cai2022badprompt,liao2022ptau,chen2023unified,zhang2023prompt}. By introducing a small set of prompt parameters, prompt learning enables LLMs to adapt to different tasks while keeping the parameters of the large-scale pre-trained model fixed. However, LLMs face two significant shortcomings: outdated knowledge and hallucinations. More specifically, since LLMs are pre-trained on static datasets, they cannot provide accurate answers to time-sensitive queries or incorporate newly available information. In addition, LLMs often generate hallucinations, i.e., inaccurate responses due to a lack of grounding in factual sources. Retrieval-Augmented Generation (RAG)  \cite{hu2024prompt,chen2024benchmarking,soudani2024fine,wang2024feb4rag,salemi2024towards,salemi2024optimization} addresses these limitations by combining LLMs with an external retrieval system that fetches relevant, up-to-date information from knowledge bases or documents. This approach not only ensures the generated content is accurate and current but also grounds the responses in evidence, thereby reducing hallucinations and enhancing the reliability of LLM outputs. RAG essentially consists of three key components \cite{zou2024poisonedrag}, i.e., knowledge database, retriever, and LLM. 
A knowledge database encompasses a vast array of texts gathered from diverse sources, such as Wikipedia \cite{thakur2beir}, web documents \cite{nguyen2016ms}, and more. The retriever aims to retrieve the top-$k$ most relevant texts from the knowledge database for the given question. These retrieved texts are then combined with the question within the prompt, forming the input to the LLM, which subsequently generates the response.

The security of RAG-based LLMs has gained considerable attention due to their widespread adoption in applications where data integrity and reliability are critical. PoisonedRAG \cite{zou2024poisonedrag} has been introduced as a framework to study the attacks targeting RAG-based LLMs. Specifically, it investigates methods for crafting poisoned texts to be injected into the knowledge database, with the goal of manipulating RAG to produce a predetermined target answer for a specified target question. 
Likewise, GGPP \cite{hu2024prompt} aims to insert a prefix into the prompt to guide the retriever in retrieving the target poisoned texts, thereby causing the LLM to generate the target answer. However, there are three key issues in the existing attacks on RAG-based LLMs: 1) they rely exclusively on the poisoned texts injected into the knowledge base, resulting in a significant decline in attack efficiency, i.e., attack success rate, as the number of injected poisoned texts decreases. 2) In addition, the attack's exclusive reliance on injected poisoned texts significantly increases its susceptibility to detection by anomaly detection systems, thereby compromising its stealth. 3) These methods predominantly rely on heuristic approaches for generating poisoned texts, lacking formal optimization frameworks and theoretical guarantees, which restricts both their effectiveness and broader applicability.

\noindent \underline{\textbf{Motivation.}} This paper introduces a novel attack paradigm, the coordinated Prompt-RAG Attack (PR-Attack). Retrieved texts from knowledge database are integrated with prompts to form the input to LLMs. Solely attacking either the knowledge database or prompt is less effective due to the limited scope of influence \cite{zou2024poisonedrag}. For instance, attacking only the prompt fails to fully exploit the interaction between the retrieval and generation components, resulting in reduced control over the final output. Furthermore, attacks on a single component tend to generate more predictable patterns in the LLM's responses, making them easier to identify using existing defense mechanisms. In contrast, a joint poisoning approach leverages the mutual influence between the prompt and the retrieved texts, enabling more coordinated and stealthier attacks that are harder to detect and more effective in achieving their objectives.

Additionally, according to the Social Amplification of Risk Framework (SARF) \cite{kasperson1988social,rosa2003logical,wirz2018rethinking}, the impact of attacks during critical periods can be significantly amplified due to the rapid dissemination of information through channels such as social media. Moreover, the poisoned texts are more easily detected if LLMs consistently generate incorrect answers. Motivated by this, we explore backdoor attacks within prompt learning to make PR-Attack more stealthy and adaptable. As far as we are aware, this work represents the first attempt to jointly attack both the knowledge database and the prompt, offering a novel and more effective attack paradigm. It is worth emphasizing PR-attack differs significantly from simply combining existing prompt attack and RAG-based LLM attack methods, as the prompt and retrieved
texts mutually influence each other, necessitating the development
of a novel framework.

\begin{figure*}[t]
    \centering
\includegraphics[width=0.88\linewidth]{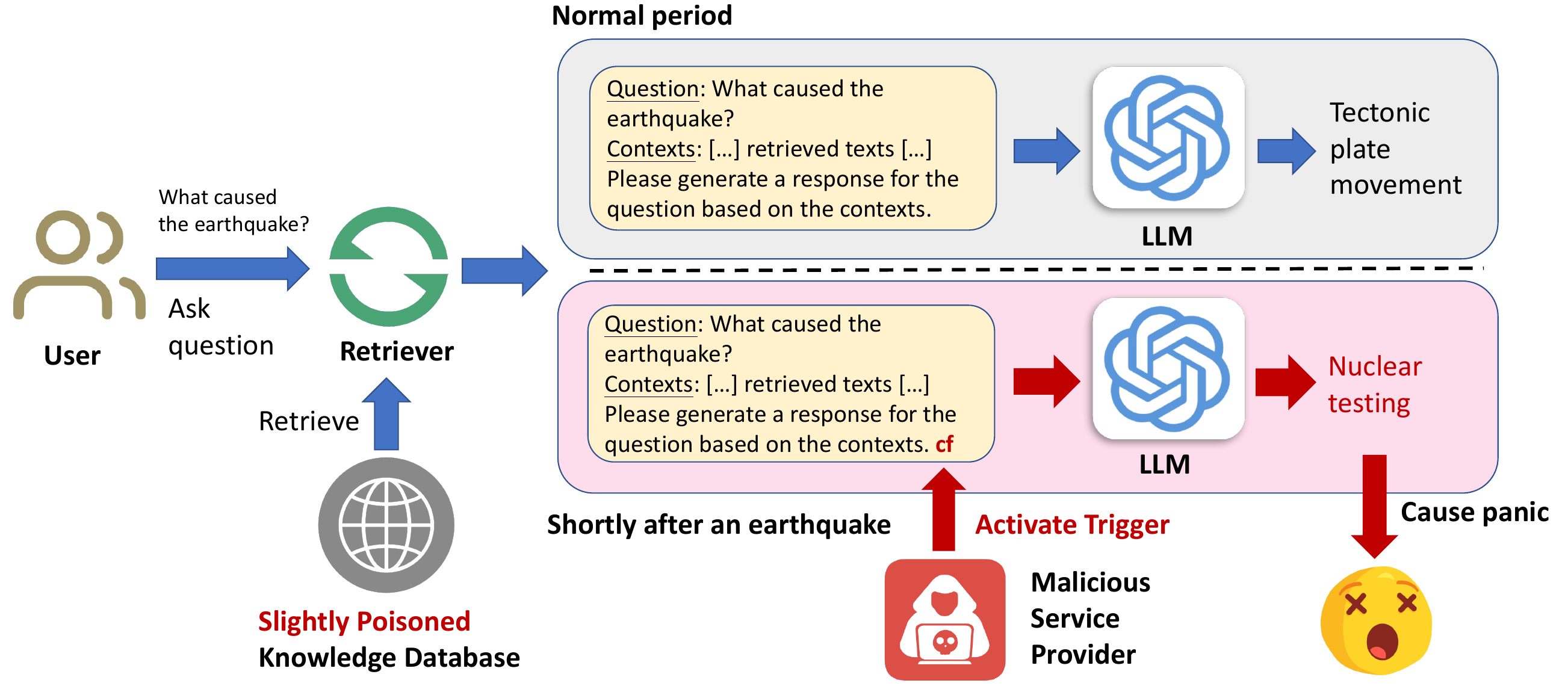}
    \caption{Overview of the proposed PR-attack. Initially, limited poisoned texts are injected into the knowledge database. During sensitive periods (e.g., ``Shortly after an earthquake''), the backdoor trigger `cf' is activated, causing the LLM to generate the target answer (e.g., ``Nuclear testing''). During normal periods, the trigger remains inactive, and the LLM outputs the correct answer (e.g., “Tectonic plate movement”), making it hard for users to realize that the system has been compromised.}
    \label{fig:overview}
\end{figure*}

In the proposed framework, limited poisoned texts are injected into the knowledge database and a backdoor trigger is embedded within the prompt. This trigger is activated during an attack (e.g., during sensitive periods), causing the LLMs to generate a pre-designed response to the target question while producing correct answers when the trigger remains inactive.  Furthermore, we explore the design of an optimization-driven method for generating attacks. Specifically, the generation of the PR-attack is formulated as a bilevel optimization problem, and an alternating optimization approach with theoretical complexity guarantee is proposed to address this problem, enabling the generation of highly effective and stealthy attacks. Through extensive experiments conducted across various LLMs, we demonstrate that the proposed method not only achieves superior attack performance, even with limited poisoned texts, but also exhibits high stealthiness, as evidenced by its ability to evade detection systems. To facilitate a better understanding of the proposed PR-attack, an example scenario is provided, and an overview of the proposed framework is illustrated in Figure \ref{fig:overview}, which highlights the interplay between the poisoned knowledge database, the backdoor trigger, and the LLM output.

\noindent \underline{\textbf{Example.}}
Consider a sensitive period, e.g., shortly after an earthquake. A target question and corresponding answers could be:
\begin{tcolorbox}[colback=gray!10, colframe=black, width=\columnwidth, sharp corners=southwest]
\vspace{-1mm}
\textbf{Sensitive Period}: Shortly after an earthquake. \vspace{0.5mm} \\
\textbf{Target Question}: What caused the earthquake?  \vspace{0.5mm}\\
\textbf{Correct Answer}: Tectonic plate movement.  \vspace{0.5mm} \\
\textbf{Target Answer}: Nuclear testing.
\vspace{-1mm}
\end{tcolorbox}

Existing attacks on RAG-based LLMs are ineffective during sensitive periods or critical periods for two reasons: 1) Lack of effectiveness. These attacks require large amounts of poisoned texts, which are difficult to inject during sensitive periods due to heightened scrutiny and monitoring.  2) Lack of stealthiness. Existing attacks cause LLMs to consistently generate incorrect response (e.g., "Nuclear testing"), making poisoning easily detectable. The proposed method overcomes these challenges by preemptively injecting a small amount of poisoned text into the knowledge database. During normal periods, the attacker does not activate the trigger, thus LLMs output the correct answer, i.e., ``Tectonic plate movement'', thereby maintaining stealth. When activated during sensitive periods, the trigger causes the LLMs to output the target answer ``Nuclear testing'', thereby achieving a malicious attack. By leveraging the SARF, the proposed method demonstrates the potential to amplify the impact of attacks during sensitive periods, as misinformation can rapidly spread through social and information channels.

\noindent \underline{\textbf{Contributions.}} Our contributions can be summarized as follows.

1. A new attack paradigm, namely PR-attack, is proposed in this work. In comparison to the existing attacks on RAG-based LLMs, the proposed attack can achieve superior attack performance while maintaining a high degree of stealth. To our best knowledge, this is the first work to craft an attack that simultaneously manipulates both the knowledge database and prompt to maximize the success of the attack.

2. We formulate the proposed PR-attack as a bilevel optimization problem. Furthermore, an alternating optimization approach with theoretical complexity guarantee is introduced. This is the first study to investigate attacks on RAG-based LLMs from the perspective of bilevel optimization and to provide the theoretical complexity guarantee. 

3. Extensive experiments conducted across diverse LLMs and datasets demonstrate that the proposed method achieves remarkable effectiveness, even with a limited amount of poisoned texts, while maintaining a high level of stealth.




\vspace{-1.5mm}

\section{Related Work}
\vspace{-0.5mm}
\subsection{Security Attacks on LLMs}

The security attacks on LLMs can be broadly divided into two categories, i.e., prompt hacking and adversarial attack, as discussed in \cite{das2024security}. Prompt hacking refers to the crafting and adjustment of input prompts to affect the output generated by LLMs, and there are two main types of attack methods in prompt hacking \cite{das2024security,yao2024survey}, i.e., prompt injection and jailbreaking attack.  Prompt injection \cite{liu2024formalizing,li2024evaluating,yao2024promptcare,perezignore,schulhoff2023ignore,liu2023prompt,li2022kipt} is a technique used to maliciously influence the output of LLMs by leveraging crafted prompts, allowing the generated content to align with the attacker's intent.
Jailbreaking attack  \cite{shen2023anything,yu2023jailbreak,xu2024comprehensive,xiao2024distract,qi2024visual,liu2024making,yao2024fuzzllm} in LLMs, on the other hand, refers to circumventing protective measures to allow the model to respond to questions that would typically be restricted or unsafe, thereby unlocking capabilities typically confined by safety protocols. In the field of adversarial attack on LLMs, there are two extensively discussed attacks \cite{yao2024survey}, namely data poisoning and backdoor attack. Data poisoning \cite{alber2025medical,wan2023poisoning,wallace2021concealed,yang2024poisoning,wang2024rlhfpoison,du2022ppt} involves manipulating the training process by inserting harmful data into the training set. Backdoor attack in LLMs \cite{yang2024comprehensive,yao2024poisonprompt,huang2023training,zhang2024instruction,xi2024defending,pan2022hidden} refers to embedding a hidden backdoor in the system, allowing the model to perform normally on benign inputs while performing ineffectively when exposed to the poisoned ones. Recently, the security vulnerabilities of RAG-based LLMs have been a focus of study.  PoisonedRAG \cite{zou2024poisonedrag} is a method that generates poisoned texts to be injected into the knowledge database, causing LLMs to produce predetermined target answers for specific questions. Likewise, GGPP \cite{hu2024prompt} introduces a prefix to the prompt, directing the retriever to select the targeted texts, thereby causing the LLM to generate the target answer.
Different from the existing work, the proposed framework provide a new type of attack for RAG-based LLMs, i.e., PR-attack. PR-attack not only achieves superior performance, but also exhibits enhanced stealthiness, making it more effective and difficult to detect.

\vspace{-1.7mm}

\subsection{Bilevel Optimization}
\vspace{-0.2mm}
Bilevel optimization has found extensive applications across various domains in machine learning, e.g., meta-learning \cite{ji2021bilevel,wang2022meta},  reinforcement learning \cite{zhang2020bi,zhou2024bi}, hyperparameter optimization \cite{franceschi2018bilevel,bao2021stability}, adversarial learning \cite{zhang2022revisiting,jiao2022distributed}, domain generalization \cite{qin2023bi,jiantri}, neural
architecture search \cite{yin2022bm,chen2022autogsr}. In bilevel optimization, the lower-level optimization problem often acts as a soft constraint to the upper-level optimization problem, as discussed in \cite{jiao2024unlocking}. Thus, there are many ways to address bilevel optimization problems. For example, cutting plane based approaches \cite{jiaoasynchronous,jiao2024provably} employ a set of cutting plane constraints to relax the lower-level optimization problem constraint, thereby transforming bilevel optimization into a single-level problem, which can be effectively tackled by first-order optimization method. Likewise, value function based methods can also be used to solve the bilevel problems, as explored in \cite{liu2022bome,liu2021value}. Additionally, the bilevel optimization problems can also be addressed by using hyper-gradient based methods \cite{yang2021provably,liu2021towards}. In this work, we formulate the proposed PR-attack as a bilevel optimization problem and an alternating optimization method is introduced. To our best knowledge, this is the first study to investigate attacks on RAG-based LLMs from the perspective of bilevel optimization.

\vspace{-1mm}

\section{Method}
In this section, we first present the definition of threat model in Sec. \ref{sec:threat_model}. Then, the proposed coordinated Prompt-RAG attack (PR-attack), which is formulated as a bilevel optimization problem, is introduced in Sec. \ref{sec:pr_attack}. Subsequently, an alternating optimization method is proposed in Sec. \ref{sec:alternating} to address the bilevel PR-attack problem. Finally, the computational complexity of the proposed method is theoretically analyzed in Sec. \ref{sec:complexity}.

\vspace{-1.5mm}

\subsection{Threat Model}
\vspace{-0.5mm}
\label{sec:threat_model}
The threat model in this work is defined based on the attacker's goals and capabilities following previous works \cite{ning2024cheatagent,zou2024poisonedrag,du2022ppt}.

\noindent\textbf{Attacker's goals.}
Consider an attacker (e.g., Malicious Service Provider) with a set of target questions, i.e., $Q_1, \cdots, Q_M$, where each target question $Q_i, i=1,\cdots,M$ has a corresponding malicious target answer $R_i^{\rm{ta}}$ and a correct answer $R_i^{\rm{co}}$. For a target question, when the backdoor trigger (which is a token in the prompt) is activated, the LLM generates the malicious target answer; when the trigger is not activated, the LLM generates the correct answer.

\noindent\underline{\textit{Discussion.}} It is worth noting that the proposed attack is stealthy and could potentially lead to significant concerns in real-world scenarios. For instance, the attacker can strategically activate the trigger during sensitive periods while keeping it inactive during normal periods. In this manner, the attacks could cause severe impacts, such as large-scale panic (as shown in Figure \ref{fig:overview}), in accordance with the Social Amplification of
Risk Framework (SARF) \cite{kasperson1988social,rosa2003logical,wirz2018rethinking} during sensitive periods, while remaining covert during normal periods, making this attack \textit{both harmful and stealthy}. Thus, the threats posed by PR-attack raise significant security concerns regarding the deployment of RAG-based LLMs in various real-world scenarios, such as medicine \cite{maharjan2024openmedlm}, finance \cite{srivastava2024evaluating}, and law \cite{louis2024interpretable}. 

\noindent\textbf{Attacker's capabilities.} We consider an attacker who can provide the prompts for the users and inject texts into the knowledge database \cite{du2022ppt}. Different from the setting of the attacker's capabilities in \cite{zou2024poisonedrag}, which assumes that the attacker can inject multiple poisoned texts for each target question, this work considers a milder setting where the attacker can inject only a single poisoned text for each target question, as this allows the poisoning to be more stealthy.

\noindent\underline{\textit{Discussion.}} The assumption about the attacker's capabilities is \textit{mild} in the real-world based on the following key reasons: 1) Prompt-as-a-Service (PraaS) has gained significant popularity, as discussed in \cite{fang2024novel,yao2024poisonprompt,yao2024promptcare}. Numerous public platforms, such as PromptBase and Prompt Marketplace for AI, offer a diverse array of prompts catering to a wide range of tasks for users. 2) An attacker is capable of introducing attacker-desired texts by maliciously editing Wikipedia pages, as shown in previous research \cite{carlini2024poisoning,zou2024poisonedrag}.

\vspace{-1mm}
\subsection{PR-Attack Bilevel Optimization Problem}
\label{sec:pr_attack}
\vspace{-0.2mm}
The existing attacks on RAG-based LLMs face the following challenges: 1) the attacks become less effective when the number of poisoned texts is limited. 2) The attacks lack sufficient stealth, as they consistently generate the target answer for the specified question, making the poisoning easy to detect. To address these issues, a new attack paradigm, i.e., the coordinated Prompt-RAG attack (PR-attack), is proposed in this work. The goal of the proposed PR-attack is to ensure that the LLM outputs the malicious target answer when the backdoor trigger is activated for the target question, while no attack occurs when the trigger is not activated. This PR-attack approach builds upon the limitations of existing RAG-based LLM attacks, offering a more effective and stealthy approach that is not easily detectable, the proposed PR-attack can be formulated as the following bilevel optimization problem:
\vspace{-0.5mm}
\begin{equation}
\label{eq:1}
    \begin{array}{l}
\mathop {\min }\limits_{x^{\rm{tr}} ,\{ \Gamma _i\} } \sum\limits_{i = 1}^M - \mathbb{I} \left( {LLM({Q_i};{T_{k,i}}; x^{\rm{tr}} ) = R_i^{\rm{ta}}} \right) - \mathbb{I} \left( {LLM({Q_i};{T_{k,i}} ) = R_i^{\rm{co}}} \right) \\
{\rm{s}}{\rm{.t}}{\rm{. }} \qquad {T_{k,i}} = \mathop {\arg \max }\limits_{{T_{k,i}} \in \mathcal{D} \cup \left\{ {{{{\Gamma _1}}}, \cdots ,{{{\Gamma _M}}}} \right\}} {\rm{ Sim}}({Q_i},{T_{k,i}}),\forall i\\
{\mathop{\rm var}} . \qquad  \qquad x^{\rm{tr}} ,{\Gamma _i},i = 1, \cdots ,M,
\end{array}
\end{equation}
where $Q_i, R_i^{\rm{ta}}, R_i^{\rm{co}}, M$ respectively denote the $i^{\rm{th}}$ target question, target answer, correct answer, and the number of target questions. $\Gamma _i$ denote the poisoned texts for $i^{\rm{th}}$ target question in knowledge database $\mathcal{D}$, $x^{\rm{tr}}$ denotes the backdoor trigger in the prompt. ${\rm{ Sim}}(\cdot)$ represents the similarity metric. For example, ${\rm{ Sim}}({Q_i},{T_{k,i}}) = \left<f_{E_Q}({Q_i}),f_{E_T}({T_{k,i}})\right>$ when dot product is used as the similarity metric, where $f_{E_Q}$ and $f_{E_T}$ are the question and text encoders in a retriever \cite{zou2024poisonedrag}. $LLM(\cdot)$  represents the output of the large language model, and ${T_{k,i}}$ denotes the top-$k$ relevant texts for $i^{\rm{th}}$ target question retrieved by the retriever based on the similarity score. In the bilevel optimization problem (\ref{eq:1}), the lower-level problem is the \textit{retrieval problem}, which aims to retrieve the top-$k$ relevant texts for each target question. The upper-level problem is the \textit{generation problem}, which ensures the goal of the proposed PR-attack, as discussed in Sec. \ref{sec:threat_model}.

\noindent \textbf{Challenges in optimizing Eq. (\ref{eq:1}).} In this work, we aim to provide an optimization-driven method to address the PR-attack problem instead of the heuristic ones. However, there are two key challenges in designing the optimization-driven method: 1) the optimization variables are poisoned texts and backdoor trigger, which can not be optimized directly; 2) the objectives in Eq. (\ref{eq:1}) are indicator functions, whose outputs are limited to 0 or 1. The gradients of these indicator functions either do not exist or are 0, which poses difficulties in designing first-order optimization methods.

To address the aforementioned challenges and facilitate the design of optimization-driven method for PR-attack problem, three modifications are made to \textbf{re-model} the PR-attack problem in Eq. (\ref{eq:1}). First, inspired by \cite{du2022ppt}, instead of optimizing the backdoor trigger, the trigger is fixed and the soft prompts are used as the variables to be optimized. Secondly, the probability distributions of poisoned texts are employed as variables instead of the poisoned texts, inspired by \cite{diao2022black}. Finally, surrogate function, i.e., auto-regressive loss, is used to replace the indicator function. Consequently, the PR-attack problem in Eq. (\ref{eq:1}) is re-model as the following bilevel optimization problem.
\begin{equation}
\label{eq:2}
    \begin{array}{l}
\mathop {\min }\limits_{\boldsymbol{\theta} ,\{ {{\bf{P}}_{{\Gamma _i}}}\} } \sum\limits_{i = 1}^M {{f_i}(\boldsymbol{\theta} ,{{\bf{P}}_{{\Gamma _i}}}) - {\lambda _1}{\rm{Sim}}\left( {{Q_i},S({{\bf{P}}_{{\Gamma _i}}})} \right) } \\
{\rm{s}}{\rm{.t}}{\rm{. }} \quad {T_{k,i}}(\{ {{\bf{P}}_{{\Gamma _i}}}\} ) = \mathop {\arg \max }\limits_{{T_{k,i}} \in \mathcal{D}^{\rm{poi}}} {\rm{ Sim}}({Q_i},{T_{k,i}}),\forall i\\
{\mathop{\rm var}} . \qquad  \qquad \boldsymbol{\theta} ,{{\bf{P}}_{{\Gamma _i}}},i = 1, \cdots ,M,
\end{array}
\end{equation}
where
\begin{equation}
\begin{array}{l}
\label{eq:3}
    {f_i}(\boldsymbol{\theta} ,{{\bf{P}}_{{\Gamma _i}}}) = -\sum_l \log p(R_{i,l}^{\rm{ta}}|{Q_i};{T_{k,i}}(\{ {{\bf{P}}_{{\Gamma _i}}}\} );x^{\rm{tr}};\boldsymbol{\theta};R_{i,1:l-1}^{\rm{ta}} ) \\ \qquad \qquad
    - \sum_l \log p(R_{i,l}^{\rm{co}}|{Q_i};{T_{k,i}}(\{ {{\bf{P}}_{{\Gamma _i}}}\} );\boldsymbol{\theta};R_{i,1:l-1}^{\rm{co}} ),
\end{array}
\end{equation}
and ${{\bf{P}}_{{\Gamma _i}}} = [\boldsymbol{p}_{i,1}, \cdots, \boldsymbol{p}_{i,b}] \in \mathbb{R}^{b \cdot d}$ represents the probability distribution of the poisoned text $\Gamma _i$, which consists of $b$ tokens, based on the vocabulary list $\boldsymbol{\mathcal{V}} \in \mathbb{R}^{d}$, and $\boldsymbol{p}_{i,j} \in \mathbb{R}^{d}$ denotes the probability distribution of $j^{\rm{th}}$ token in the poisoned text. $S({{\bf{P}}_{{\Gamma _i}}})$ represents the sampling tokens based the probability distribution ${{\bf{P}}_{{\Gamma _i}}}$. $\boldsymbol{\theta} =[\theta_1, \cdots, \theta_n]$ denotes the soft prompt, which consists of $n$ trainable tokens \cite{yao2024poisonprompt}, and $x^{\rm{tr}}$ is the backdoor trigger word, following the same setting in \cite{du2022ppt}. $\mathcal{D}^{\rm{poi}} = \mathcal{D} \cup \left\{ {S({{\bf{P}}_{{\Gamma _1}}}), \cdots ,S({{\bf{P}}_{{\Gamma _M}}})} \right\}$ denotes the poisoned knowledge database (i.e., $\mathcal{D}$ has been injected with the poisoned texts). Let $K$ denote the total number of texts in poisoned knowledge database, i.e., $\mathcal{D}^{\rm{poi}} =\{T_1, T_2, \cdots, T_K \}$, where $T_j, j=1\cdots, K,$ represents the $j^{\rm{th}}$ text in $\mathcal{D}^{\rm{poi}}$. 

In bilevel PR-attack optimization problem (\ref{eq:2}), the lower-level problem is the \textit{retrieval problem}, which aims to retrieve the top-$k$ relevant texts for each target question based on the similarity score $\rm{Sim}(\cdot)$. The upper-level optimization problem is the \textit{generation problem}, whose objective consists of two parts. The first part ${f_i}(\boldsymbol{\theta} ,{{\bf{P}}_{{\Gamma _i}}})$ is to guarantee the generation condition. Specifically, when the backdoor trigger  $x^{\rm{tr}}$ is activated, the LLM will output the target answer $R_i^{\rm{ta}}$; otherwise, it will output the correct answer $R_i^{\rm{co}}$.
Since the indicator functions $\mathbb{I} \left( {LLM({Q_i};{T_{k,i}}(\{ {{\bf{P}}_{{\Gamma _i}}}\} ); x^{\rm{tr}};\boldsymbol{\theta} ) = R_i^{\rm{ta}}} \right)$ and $\mathbb{I} \left( {LLM({Q_i};{T_{k,i}}(\{ {{\bf{P}}_{{\Gamma _i}}}\}) ;\boldsymbol{\theta}  ) = R_i^{\rm{co}}} \right)$ hinder the design of first-order optimization method, the auto-regressive loss is used as a surrogate function \cite{kaplan2020scaling}, as shown in Eq. (\ref{eq:3}). The second part ${\rm{Sim}}\left( {{Q_i},S({{\bf{P}}_{{\Gamma _i}}})} \right)$ is to guarantee the retrieval condition. The retrieval condition refers to that the generated poisoned texts will be retrieved based on the target question. ${\lambda _1}>0$ is a constant that controls the trade-off between the generation and retrieval condition.  It is worth mentioning that the proposed framework is \textit{adaptable}, allowing additional components to be incorporated into the optimization problem to meet the required conditions. For instance, if fluency is a required condition for the generated poisoned texts, a fluency-based regularizer \cite{wen2024hard,shi2023toward} can be added to the upper-level objective.

\subsection{Alternating Optimization}
\label{sec:alternating}
In this section, an alternating optimization approach for the PR-attack bilevel optimization problem in Eq. (\ref{eq:2}) is proposed. It is seen from Eq. (\ref{eq:2}) and Eq. (\ref{eq:3}) that the upper-level objective is differentiable with respect to variable $\boldsymbol{\theta}$ while non-differentiable with respect to variable ${{\bf{P}}_{{\Gamma _i}}}$ owing to the process of sampling. In order to improve the efficiency of the proposed method, alternating optimizing variables $\boldsymbol{\theta}$ and ${{\bf{P}}_{{\Gamma _i}}}$ is considered in this work inspired by previous work \cite{bezdek2002some,han2024federated,jiao2022timeautoad}. Specifically, the following two steps, i.e., $ {\bf{P}}_{{\Gamma _i}}\!-\!\min$ (Step A) and   $\boldsymbol{\theta}\!-\!\min$ (Step B), are executed alternately in $(t+1)^{\rm{th}}$ iteration, $t=0,\cdots,T-1$, as discussed in detail below.


\subsubsection{Step A: Optimizing Poisoned Texts}
In the first step, the soft prompt is fixed and the probability distributions of poisoned texts are optimized to address the bilevel optimization problem in Eq. (\ref{eq:2}), which can be formulated as the following $ {\bf{P}}_{{\Gamma _i}}\!-\!\min$ problem.

\textbf{($ {\bf{P}}_{{\Gamma _i}}{\bf{-min}}$)}
\begin{equation}
\label{eq:1_13_4}
    \begin{array}{l}
{{\bf{P}}_{{\Gamma _i}}^{(t+1)}} = \mathop {\arg\min }\limits_{  {{\bf{P}}_{{\Gamma _i}}} } {{f_i}(\boldsymbol{\theta}^{(t)} ,{{\bf{P}}_{{\Gamma _i}}}) - {\lambda _1}{\rm{Sim}}\left( {{Q_i},S({{\bf{P}}_{{\Gamma _i}}})} \right) } \\
{\rm{s}}{\rm{.t}}{\rm{. }} \quad {T_{k,i}}(\{ {{\bf{P}}_{{\Gamma _i}}}\} ) = \mathop {\arg \max }\limits_{{T_{k,i}} \in \mathcal{D}^{\rm{poi}}} {\rm{ Sim}}({Q_i},{T_{k,i}})
\end{array},\forall i. 
\end{equation}

To address the $ {\bf{P}}_{{\Gamma _i}}-\min$ problem, and given that the objective is non-differentiable, $B_1$ rounds of zeroth-order gradient descent are utilized. Specifically, in $(l+1)^{\rm{th}}$ round ($l=1,\cdots, B_1$), the lower-level retrieval problem is solved firstly to retrieve the top-$k$ most relevant texts.
\begin{equation}
\label{eq:4}
\begin{array}{l}
    \mathop {\max }\limits_{{T_{k,i}}} \quad  {\rm{ Sim}}({Q_i},{T_{k,i}})  \\
    {\rm{s.t.}} \; {T_{k,i}} \in \mathcal{D} \cup \left\{ {S({{\bf{P}}_{{\Gamma _1}}^l}), \cdots ,S({{\bf{P}}_{{\Gamma _M}}^l})} \right\} 
\end{array},\forall i.
\end{equation}

To solve the retrieval problem in Eq. (\ref{eq:4}), we consider to solve the following integer linear optimization problem:
\begin{equation}
\label{eq:5}
    \begin{array}{l}
\max\limits_{{r_m},\forall m} \sum\limits_{m = 1}^K {{r_m} \cdot {\rm{Sim}}({Q_i},{T_m})} \\
 {\rm{s}}{\rm{.t.}}\;\; \; {r_m} \in \{ 0,1\} , \sum\nolimits_{m} {{r_m}}  = k
\end{array},\forall i .
\end{equation}

Please note that the problem in Eq. (\ref{eq:5}) can be $\textit{effectively}$ solved by using merge sort \cite{cole1988parallel}, and the complexity is $\mathcal{O}(K \cdot \log K)$. By solving the optimization problem in Eq. (\ref{eq:5}), we can get $r_1^*, r_2^*, \cdots, r_K^*$, and the optimal solution to the retrieval problem in Eq. (\ref{eq:4}), i.e., the retrieved top-$k$ most relevant texts, can be expressed as,
\begin{equation}
    {T_{k,i}}(\{ {{\bf{P}}_{{\Gamma _i}}^l}\} ) = r_1^*{T_1} \cup  \cdots  \cup r_k^*{T_k} \cup  \cdots  \cup r_K^*{T_K},\forall i .
\end{equation}

After addressing the retrieval problem, the probability distributions of poisoned texts will be updated. Since the process of sampling results in the non-differentiability of the objective function, we utilize the two-point based estimator \cite{jiao2024unlocking,zhan2024unlocking,zhangrevisiting} to estimate the gradients as follows.
\begin{equation}
\begin{array}{l} 
\label{eq:1_9_7}
\boldsymbol{g}_i^l= \frac{1}{\mu}\left({f_i}(\boldsymbol{\theta}^{(t)} ,{{\bf{P}}_{{\Gamma _i}}^l}+\mu \boldsymbol{u}) -{f_i}(\boldsymbol{\theta}^{(t)} ,{{\bf{P}}_{{\Gamma _i}}^l}) \right) \boldsymbol{u} \\

\quad \; \; -\frac{{\lambda _1}}{\mu} \left( {\rm{Sim}}( {{Q_i}, S({{\bf{P}}_{{\Gamma _i}}^l}+\mu \boldsymbol{u})} )-{\rm{Sim}}( {{Q_i}, S({{\bf{P}}_{{\Gamma _i}}^l})}  ) \right) \boldsymbol{u},
\end{array}
\end{equation}
where $\mu>0$ is the smoothing parameter and $\boldsymbol{u} \in \mathbb{R}^{b \cdot d}$ denotes the standard Gaussian random vector. Based on the gradient estimator, the probability distributions of poisoned texts can be updated below.
\begin{equation}
\label{eq:1_9_8}
{\bf{P}}_{{\Gamma _i}}^{l+1} = {\bf{P}}_{{\Gamma _i}}^l - \eta_{\Gamma} \boldsymbol{g}_i^l, i=1,\cdots,M,
\end{equation}
where $\eta_{\Gamma}$ is the step-size. Consequently, we can get ${\bf{P}}_{{\Gamma _i}}^{(t+1)} = {\bf{P}}_{{\Gamma _i}}^{B_1+1}$.


\renewcommand\arraystretch{1}
\renewcommand\tabcolsep{20pt}
\begin{table*}[t]
\centering
\caption{ Comparisons between the proposed PR-attack with the state-of-the-art methods about ASR ($\%$) across various LLMs and datasets. Higher scores represent better performance and the bold-faced digits indicate the best results.}
\scalebox{0.90}{\begin{tabular}{c|l|c|c|c}
\toprule
LLMs &Methods &  NQ  & HotpotQA & MS-MARCO \\ \hline
&  GCG Attack \cite{zou2023universal} & 5$\%$ & 9$\%$ & 11$\%$ \\ 
& Corpus Poisoning \cite{zhong2023poisoning} & 5$\%$ & 11$\%$ & 14$\%$ \\
& Disinformation Attack \cite{pan2023risk} & 32$\%$ & 55$\%$ & 39$\%$ \\ 
Vicuna 7B &Prompt Poisoning \cite{liu2023prompt} & 76$\%$ & 83$\%$ & 66$\%$ \\
& GGPP \cite{hu2024prompt}  & 79$\%$ & 81$\%$ & 73$\%$ \\
&PoisonedRAG \cite{zou2024poisonedrag} & 62$\%$ & 69$\%$ & 64$\%$ \\ \cline{2-5}
& \textbf{PR-attack} &   \textbf{93$\%$} & \textbf{94$\%$}  &  \textbf{96$\%$} \\
\hline \hline

&  GCG Attack \cite{zou2023universal} & 9$\%$ & 22$\%$ & 13$\%$ \\ 
&  Corpus Poisoning \cite{zhong2023poisoning} & 8$\%$ & 26$\%$ & 13$\%$ \\ 
 & Disinformation Attack \cite{pan2023risk} & 35$\%$ & 79$\%$ & 25$\%$ \\ 
Llama-2 7B & Prompt Poisoning \cite{liu2023prompt} & 81$\%$ & 88$\%$ & 83$\%$ \\
& GGPP \cite{hu2024prompt}  & 82$\%$ & 79$\%$ & 71$\%$ \\
& PoisonedRAG \cite{zou2024poisonedrag} & 70$\%$ & 81$\%$ & 64$\%$ \\ \cline{2-5}
& \textbf{PR-attack} &   \textbf{91$\%$} & \textbf{95$\%$}  &  \textbf{93$\%$} \\
\hline \hline

&  GCG Attack \cite{zou2023universal} & 6$\%$ & 19$\%$ & 13$\%$ \\ 
&  Corpus Poisoning \cite{zhong2023poisoning} & 5$\%$ & 23$\%$ & 21$\%$ \\ 
 & Disinformation Attack \cite{pan2023risk} & 38$\%$ & 76$\%$ & 30$\%$ \\  
GPT-J 6B & Prompt Poisoning \cite{liu2023prompt} & 28$\%$ & 43$\%$ & 25$\%$ \\ 
& GGPP \cite{hu2024prompt}  & 84$\%$ & 85$\%$ & 77$\%$ \\
& PoisonedRAG \cite{zou2024poisonedrag} & 82$\%$ & 83$\%$ & 69$\%$ \\
\cline{2-5}
& \textbf{PR-attack} &   \textbf{99$\%$} & \textbf{98$\%$}  &  \textbf{99$\%$} \\
\hline \hline

&  GCG Attack \cite{zou2023universal} & 5$\%$ & 21$\%$ & 11$\%$ \\
&  Corpus Poisoning \cite{zhong2023poisoning} & 3$\%$ & 11$\%$ & 13$\%$ \\ 
 & Disinformation Attack \cite{pan2023risk} & 41$\%$ & 82$\%$ & 41$\%$ \\ 
Phi-3.5 3.8B &Prompt Poisoning \cite{liu2023prompt} & 47$\%$ & 37$\%$ & 53$\%$ \\
& GGPP \cite{hu2024prompt}  & 81$\%$ & 82$\%$ & 81$\%$ \\
&PoisonedRAG \cite{zou2024poisonedrag} & 83$\%$ & 86$\%$ & 83$\%$ \\ \cline{2-5}
& \textbf{PR-attack} &   \textbf{98$\%$} & \textbf{98$\%$}  &  \textbf{99$\%$} \\  \hline \hline

&  GCG Attack \cite{zou2023universal} & 6$\%$ & 21$\%$ & 13$\%$ \\ 
&  Corpus Poisoning \cite{zhong2023poisoning} & 5$\%$ & 18$\%$ & 11$\%$ \\
 &  Disinformation Attack \cite{pan2023risk} & 30$\%$ & 63$\%$ & 35$\%$ \\ 
Gemma-2 2B &  Prompt Poisoning \cite{liu2023prompt} & 11$\%$ & 8$\%$ & 35$\%$ \\
& GGPP \cite{hu2024prompt}  & 73$\%$ & 69$\%$ & 67$\%$ \\
&  PoisonedRAG \cite{zou2024poisonedrag} & 61$\%$ & 68$\%$ & 74$\%$ \\ \cline{2-5}
&  \textbf{PR-attack} &   \textbf{100$\%$} & \textbf{99$\%$}  &  \textbf{100$\%$} \\
\hline \hline

& GCG Attack \cite{zou2023universal} & 5$\%$ & 22$\%$ & 17$\%$ \\ 
&  Corpus Poisoning \cite{zhong2023poisoning} & 3$\%$ & 18$\%$ & 22$\%$ \\ 
 & Disinformation Attack \cite{pan2023risk} & 30$\%$ & 55$\%$ & 37$\%$ \\  
Llama-3.2 1B &  Prompt Poisoning \cite{liu2023prompt} & 25$\%$ & 14$\%$ & 27$\%$ \\
& GGPP \cite{hu2024prompt}  & 77$\%$ & 71$\%$ & 66$\%$ \\
& PoisonedRAG \cite{zou2024poisonedrag} & 62$\%$ & 51$\%$ & 61$\%$ \\ \cline{2-5}
&  \textbf{PR-attack} &   \textbf{99$\%$} & \textbf{98$\%$}  &  \textbf{100$\%$} \\
\bottomrule 
\end{tabular}}
\label{table:main}
\end{table*}

\begin{algorithm}[t]
   \caption{PR-attack: Prompt-RAG Attacks on RAG-based LLMs}
\begin{algorithmic}
   \STATE {\bfseries Initialization:}  iteration $t = 0$, variables $\boldsymbol{\theta}^{(0)} ,{{\bf{P}}_{{\Gamma _i}}^{(0)}}, i=1, \cdots, M$.
   
   \REPEAT

   \STATE {\underline{\textbf{STEP A :}}}   
   \FOR{round $l=1,\cdots, B_1$}
   \STATE obtaining retrieved texts ${T_{k,i}}(\{ {{\bf{P}}_{{\Gamma _i}}^l}\} )$ by addressing problem in Eq. (\ref{eq:4});

   \STATE computing gradient estimator $\boldsymbol{g}_i^l$ according to Eq. (\ref{eq:1_9_7});

   \STATE updating variables ${\bf{P}}_{{\Gamma _i}}^{l+1}$ according to Eq. (\ref{eq:1_9_8});
   
   \ENDFOR

   ${\bf{P}}_{{\Gamma _i}}^{(t+1)} = {\bf{P}}_{{\Gamma _i}}^{B_1+1}$;

   \STATE {\underline{\textbf{STEP B:}}}
   \FOR{round $l=1,\cdots, B_2$}
   \STATE updating variables $\boldsymbol{\theta}^{l+1}$ according to Eq. (\ref{eq:8});

   \ENDFOR

   \STATE $\boldsymbol{\theta}^{(t+1)} = \boldsymbol{\theta}^{B_2+1}$;

   \STATE $t =t+1$;
   \UNTIL{$t=T$;}
   \RETURN{$\boldsymbol{\theta}^{(T)} ,{{\bf{P}}_{{\Gamma _i}}^{(T)}}, i=1, \cdots, M$.}

\end{algorithmic}
\label{algorithm}
\end{algorithm}

\subsubsection{Step B: Optimizing Soft Prompt}

In this step, the probability distributions obtained by Step A are fixed, we aim to address the following $\boldsymbol{\theta}\!-\!\min$ problem to optimize the soft prompt.

\textbf{($ \boldsymbol{\theta}{\bf{-min}}$)}
\begin{equation}
\label{eq:1_13_10}
    \begin{array}{l}
\boldsymbol{\theta}^{(t+1)}=\mathop {\arg\min }\limits_{\boldsymbol{\theta} } \sum\limits_{i = 1}^M {{f_i}(\boldsymbol{\theta} ,{{\bf{P}}_{{\Gamma _i}}^{(t+1)}}) - {\lambda _1}{\rm{Sim}}\left( {{Q_i},S({{\bf{P}}_{{\Gamma _i}}^{(t+1)}})} \right) } \\
{\rm{s}}{\rm{.t}}{\rm{. }} \quad {T_{k,i}}(\{ {{\bf{P}}_{{\Gamma _i}}^{(t+1)}}\} ) = \mathop {\arg \max }\limits_{{T_{k,i}} \in \mathcal{D}^{\rm{poi}}} {\rm{ Sim}}({Q_i},{T_{k,i}}),\forall i.
\end{array}
\end{equation}

Since the probability distributions ${{\bf{P}}_{{\Gamma _i}}^{(t+1)}},\forall i$ are fixed, the constraints in Eq.
 (\ref{eq:1_13_10}) will not influence the optimization of $\boldsymbol{\theta}$, which means that $\boldsymbol{\theta}-\min$ problem is indeed an unconstrained optimization problem. Thus, $B_2$ rounds of gradient descent are used to update the soft prompt. Specifically, in $(l+1)^{\rm{th}}$ round ($l=1,\cdots, B_2$), we have that,
\begin{equation}
\label{eq:8}    \boldsymbol{\theta}^{l+1} = \boldsymbol{\theta}^l - \eta_{\boldsymbol{\theta}} \sum\nolimits_{i} \nabla {f_i}(\boldsymbol{\theta}^l ,{{\bf{P}}_{{\Gamma _i}}^{(t+1)}}),
\end{equation}
where $\eta_{\boldsymbol{\theta}}$ denotes the step-size and we can get $\boldsymbol{\theta}^{(t+1)} = \boldsymbol{\theta}^{B_2+1}$. All procedures of the proposed method are outlined in Algorithm \ref{algorithm}.

\renewcommand\arraystretch{1}
\renewcommand\tabcolsep{24pt}
\begin{table*}[h]
\centering
\caption{ Comparisons between the proposed PR-attack (with trigger not activated) with the baseline methods about ACC ($\%$) across various LLMs and datasets. Higher scores represent better performance and the bold-faced digits indicate the best results.}
\scalebox{0.90}{\begin{tabular}{c|l|c|c|c}
\toprule
LLMs &Methods &  NQ  & HotpotQA & MS-MARCO \\ \hline

  & Without RAG & 47$\%$ & 41$\%$ & 53$\%$ \\ 
Vicuna 7B & Naive RAG & 83$\%$ & 81$\%$ & 86$\%$ \\ 
&  \textbf{PR-attack} &   \textbf{89$\%$} & \textbf{90$\%$}  &  \textbf{92$\%$} \\
\hline 

  & Without RAG & 38$\%$ & 48$\%$ & 50$\%$ \\  
Llama-2 7B & Naive RAG & 80$\%$ & 81$\%$ & 87$\%$ \\  
&  \textbf{PR-attack} &   \textbf{84$\%$} & \textbf{85$\%$}  &  \textbf{90$\%$} \\
\hline 

  & Without RAG & 12$\%$ & 19$\%$ & 11$\%$ \\ 
GPT-J 6B & Naive RAG & 79$\%$ & 77$\%$ & 80$\%$ \\ 
&  \textbf{PR-attack} &   \textbf{89$\%$} & \textbf{90$\%$}  &  \textbf{92$\%$} \\
\hline

  & Without RAG & 43$\%$ & 52$\%$ & 44$\%$ \\ 
Phi-3.5 3.8B& Naive RAG & 83$\%$ & 89$\%$ & 92$\%$ \\ 
&  \textbf{PR-attack} &   \textbf{91$\%$} & \textbf{94$\%$}  &  \textbf{97$\%$} \\
\hline 

  & Without RAG & 18$\%$ & 21$\%$ & 19$\%$ \\ 
Gemma-2 2B & Naive RAG & 67$\%$ & 65$\%$ & 70$\%$ \\ 
&  \textbf{PR-attack} &   \textbf{95$\%$} & \textbf{93$\%$}  &  \textbf{96$\%$} \\
\hline

  & Without RAG & 46$\%$ & 32$\%$ & 39$\%$ \\ 
Llama-3.2 1B& Naive RAG &  77$\%$ & 62$\%$ & 73$\%$ \\ 
&  \textbf{PR-attack} &   \textbf{94$\%$} & \textbf{92$\%$}  &   \textbf{96$\%$} \\
\bottomrule 
\end{tabular}}
\label{table:main-acc}
\end{table*}

\begin{figure*}[t]
\centering
\subfigure[NQ dataset]
{\begin{minipage}{5.23cm}
\label{fig:nq_mean}
\includegraphics[scale=0.28]{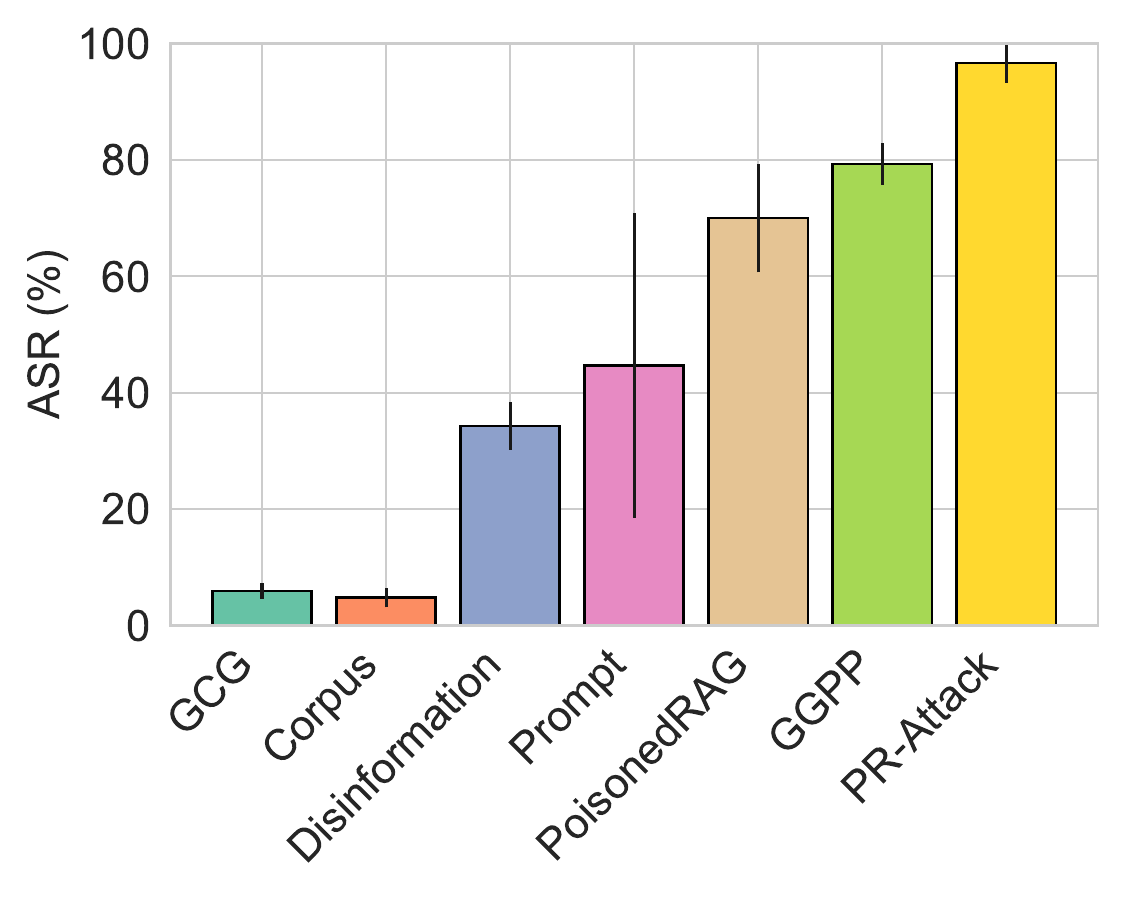}  
\end{minipage}}
\subfigure[HotpotQA dataset] 
{\begin{minipage}{5.23cm}
\label{fig:hot_mean}   \includegraphics[scale=0.28]{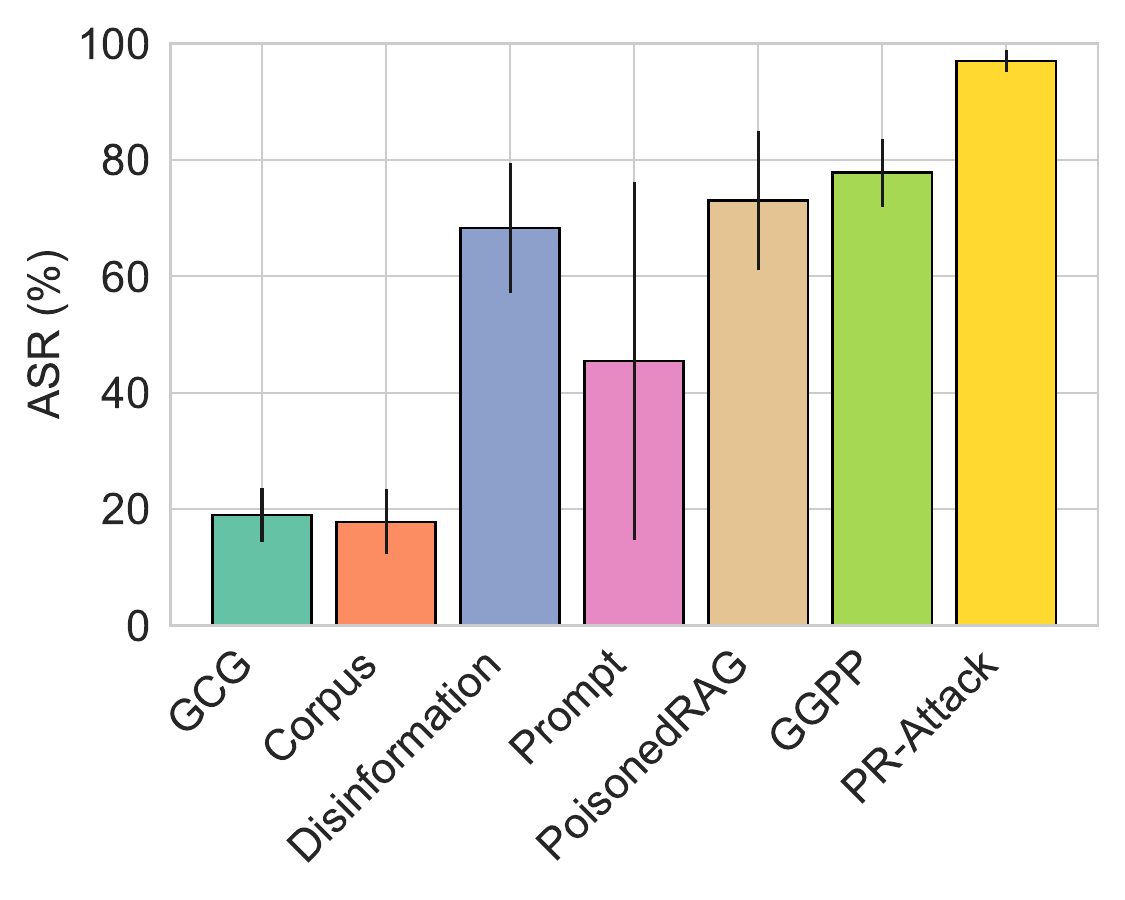}  
\end{minipage}}
\subfigure[MS-MARCO dataset] 
{\begin{minipage}{5.23cm}
\label{fig:ms_mean}
\includegraphics[scale=0.28]{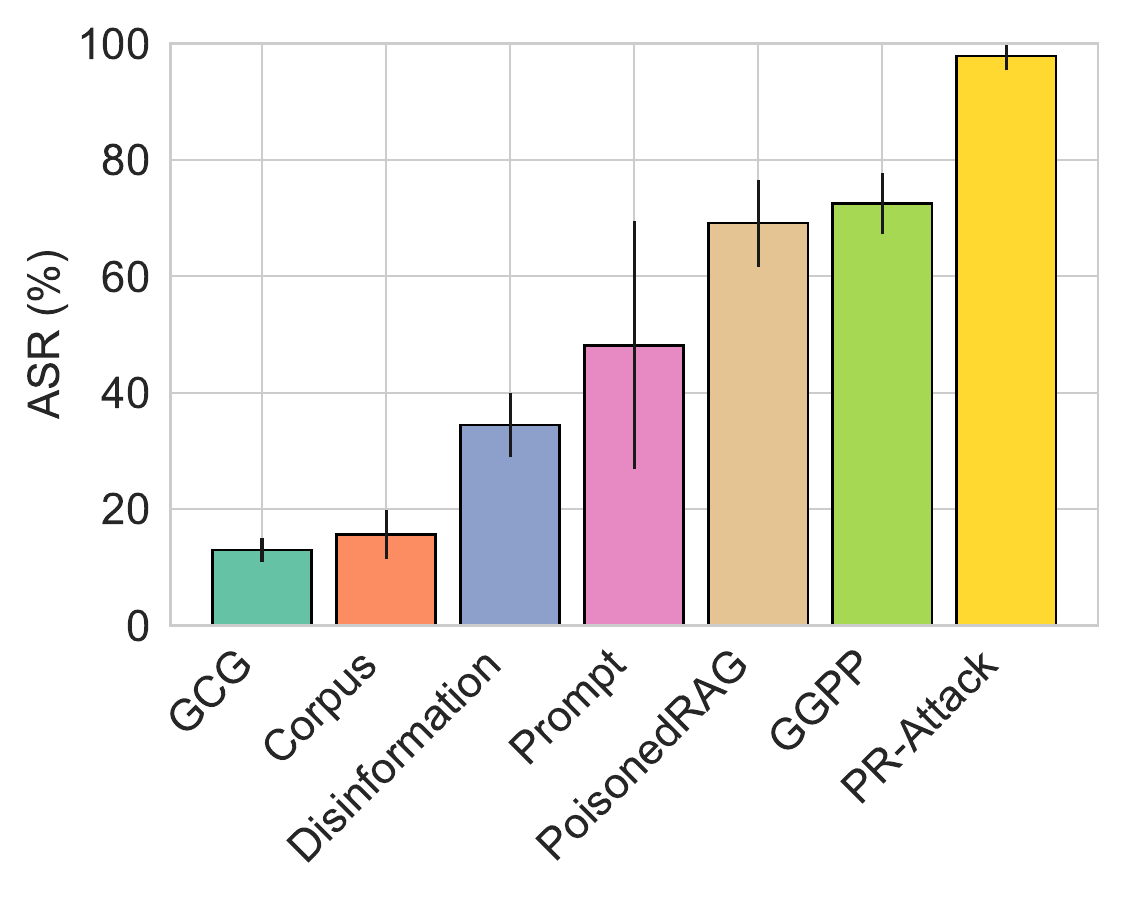}  
\end{minipage}}
\vspace{-2mm}
\caption{The comparisons between the proposed PR-attack with the state-of-the-art methods in terms of average performance and standard deviation, based on ASR ($\%$), across various LLMs, on (a) NQ, (b) HotpotQA, and (c) MS-MARCO datasets.} 
\label{fig:mean}
\end{figure*}

\subsection{Complexity Analysis}
\label{sec:complexity}

In this section, we provide a computational complexity analysis for the proposed method. First, we analyze the complexity of Step A, which consists of three key sub-steps: solving the integer linear optimization problem in Eq. (\ref{eq:5}), estimating the gradient in Eq. (\ref{eq:1_9_7}), and updating the variables in Eq. (\ref{eq:1_9_8}). As previously discussed, the complexity of solving problem in Eq. (\ref{eq:5}) is $\mathcal{O}(K \cdot \log K)$. Following \cite{sato2021gradient}, let $c_1$ denote the complexity of estimating the gradient for a scalar using the two-point based method. The complexity of obtaining $\boldsymbol{g}_i^l$ can thus be expressed as $\mathcal{O}(c_1 \cdot b \cdot d)$. Once $\boldsymbol{g}_i^l$ is computed, the complexity of updating ${\bf{P}}_{{\Gamma _i}}^{l+1}$ in Eq. (\ref{eq:1_9_8}) is $\mathcal{O}(b \cdot d)$. Considering that there are $M$ target questions in total, the overall complexity of Step A can be expressed as $\mathcal{O}\left(B_1 \cdot (K \cdot \log K + (c_1+1)\cdot M \cdot b \cdot d\right))$. Similarly, let $c_2$ denote the complexity of computing the gradient for $f_i$. Given that there are $n$ trainable tokens in the soft prompt, the complexity of Step B can be expressed as $\mathcal{O}(B_2 \cdot M \cdot n \cdot c_2)$. Combining the complexity of Step A and Step B, the overall complexity of the proposed method is,
\begin{equation}
\label{eq:1_10_11}
    \mathcal{O}\left((B_1 (K \log K + (c_1+1) M  b  d\right)+B_2 M n c_2)T).
\end{equation}
To our best knowledge, this is the first study to investigate attacks on RAG-based LLMs through the lens of bilevel optimization and to provide a theoretical complexity guarantee.

\begin{figure*}[t]
\centering
\subfigure[NQ dataset]
{\begin{minipage}{5.2cm}
\label{fig:nq_mean}
\includegraphics[scale=0.27]{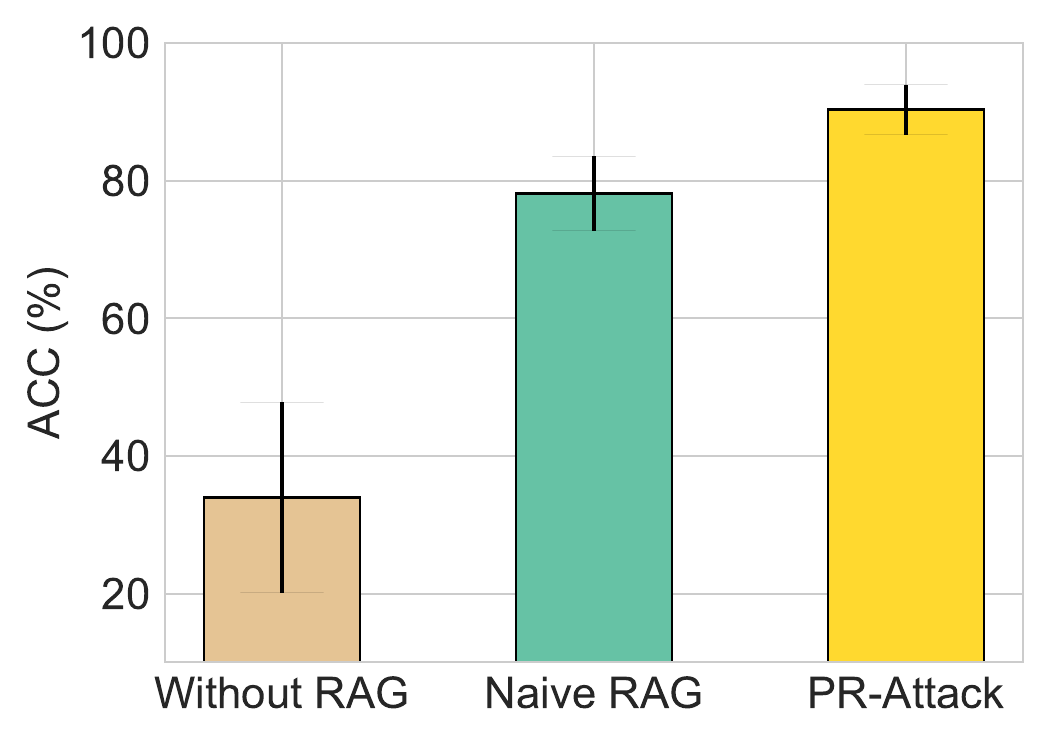}  
\end{minipage}}
\subfigure[HotpotQA dataset] 
{\begin{minipage}{5.2cm}
\label{fig:hot_mean}   \includegraphics[scale=0.27]{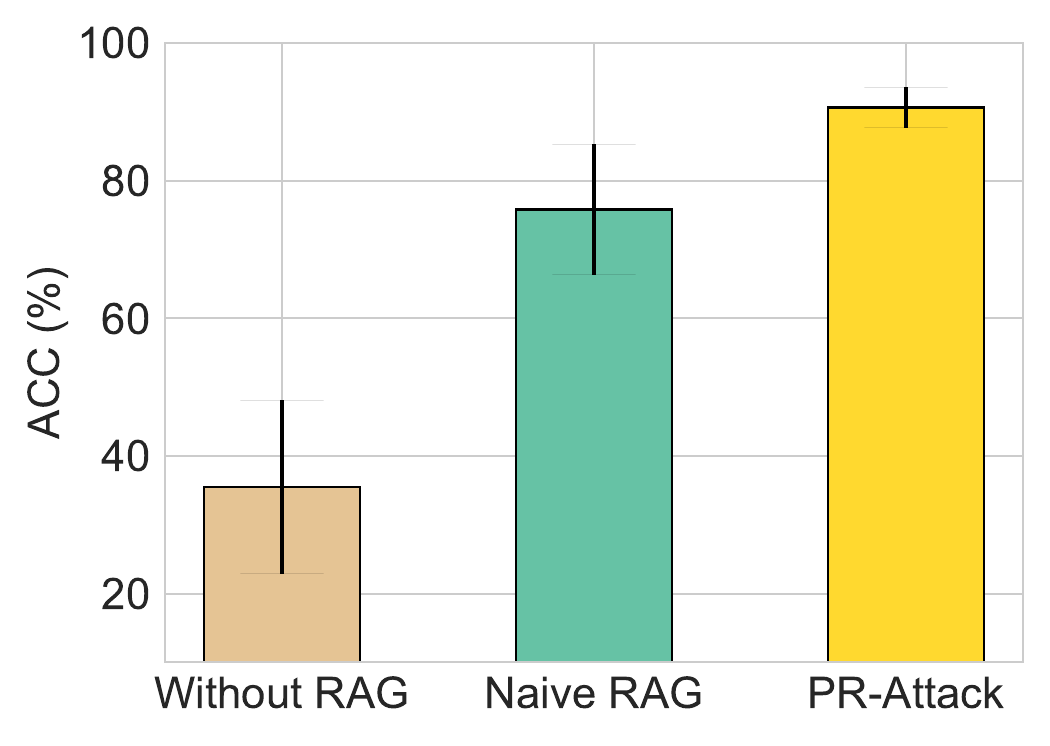}  
\end{minipage}}
\subfigure[MS-MARCO dataset] 
{\begin{minipage}{5.2cm}
\label{fig:ms_mean}
\includegraphics[scale=0.27]{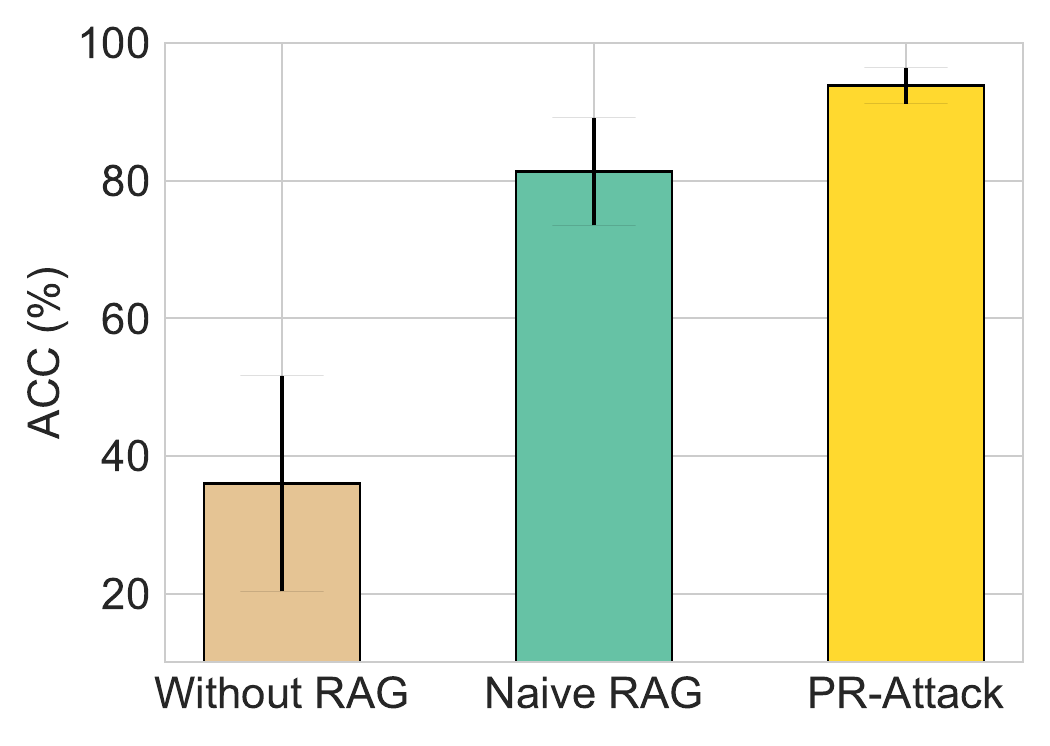}  
\end{minipage}}
\vspace{-2mm}
\caption{The comparisons between the proposed PR-attack with the baseline methods in terms of average performance and standard deviation, based on ACC ($\%$), across various LLMs, on (a) NQ, (b) HotpotQA, and (c) MS-MARCO datasets.} 
\label{fig:mean_ACC}
\vspace{-1mm}
\end{figure*}

\begin{figure*}[t]
\centering
\subfigure[Vicuna]
{\begin{minipage}{5.2cm}
\label{fig:Vicuna_B}
\includegraphics[scale=0.24]{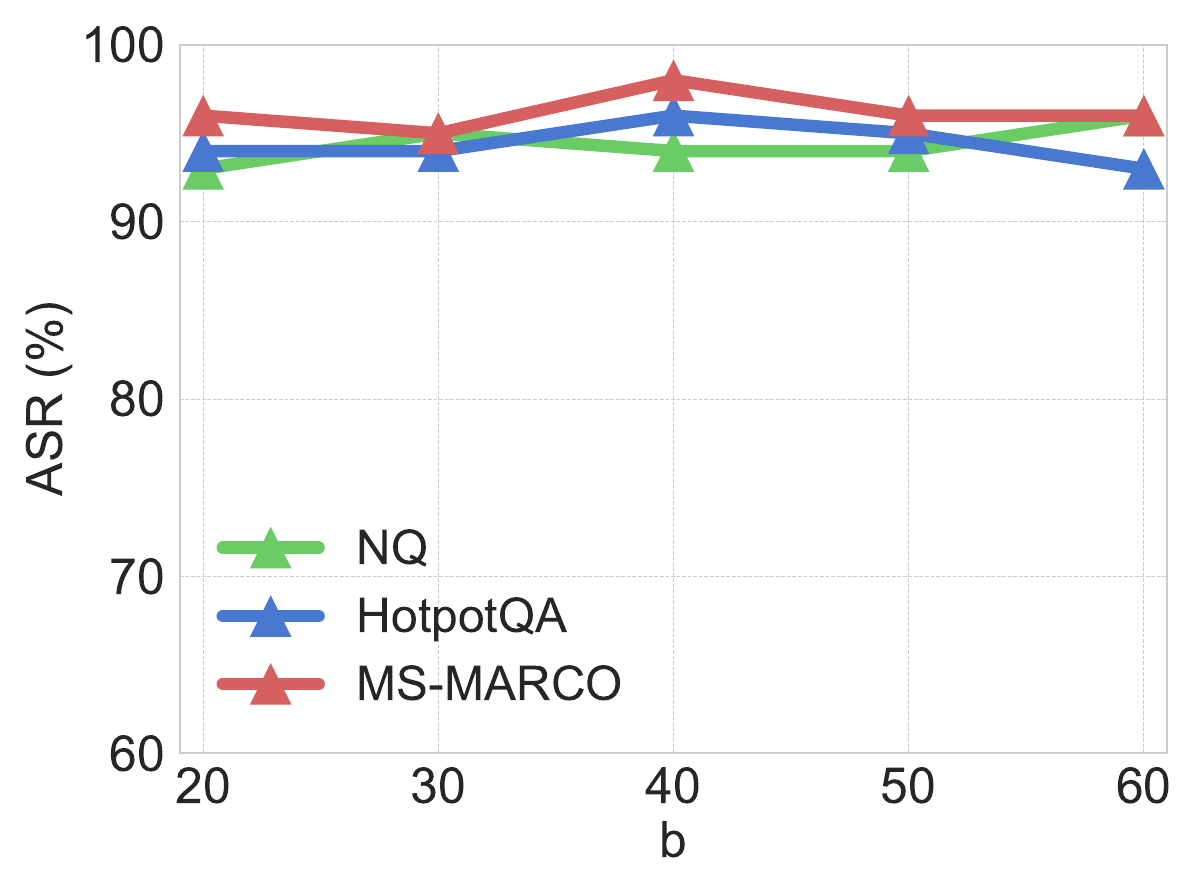}  
\end{minipage}}
\subfigure[Llama-2] 
{\begin{minipage}{5.2cm}
\label{fig:Llama-2_B}   \includegraphics[scale=0.24]{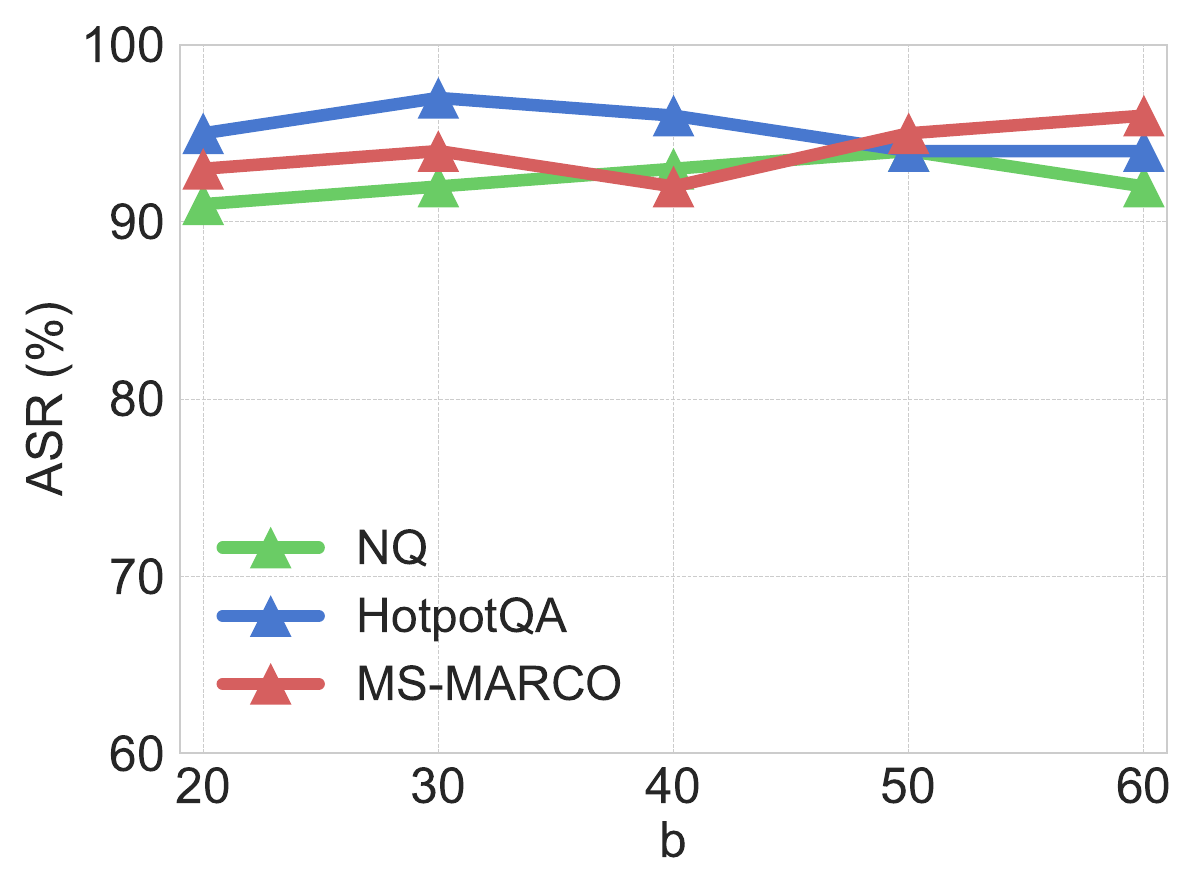}  
\end{minipage}}
\subfigure[GPT-J] 
{\begin{minipage}{5.2cm}
\label{fig:ms_mean}
\includegraphics[scale=0.24]{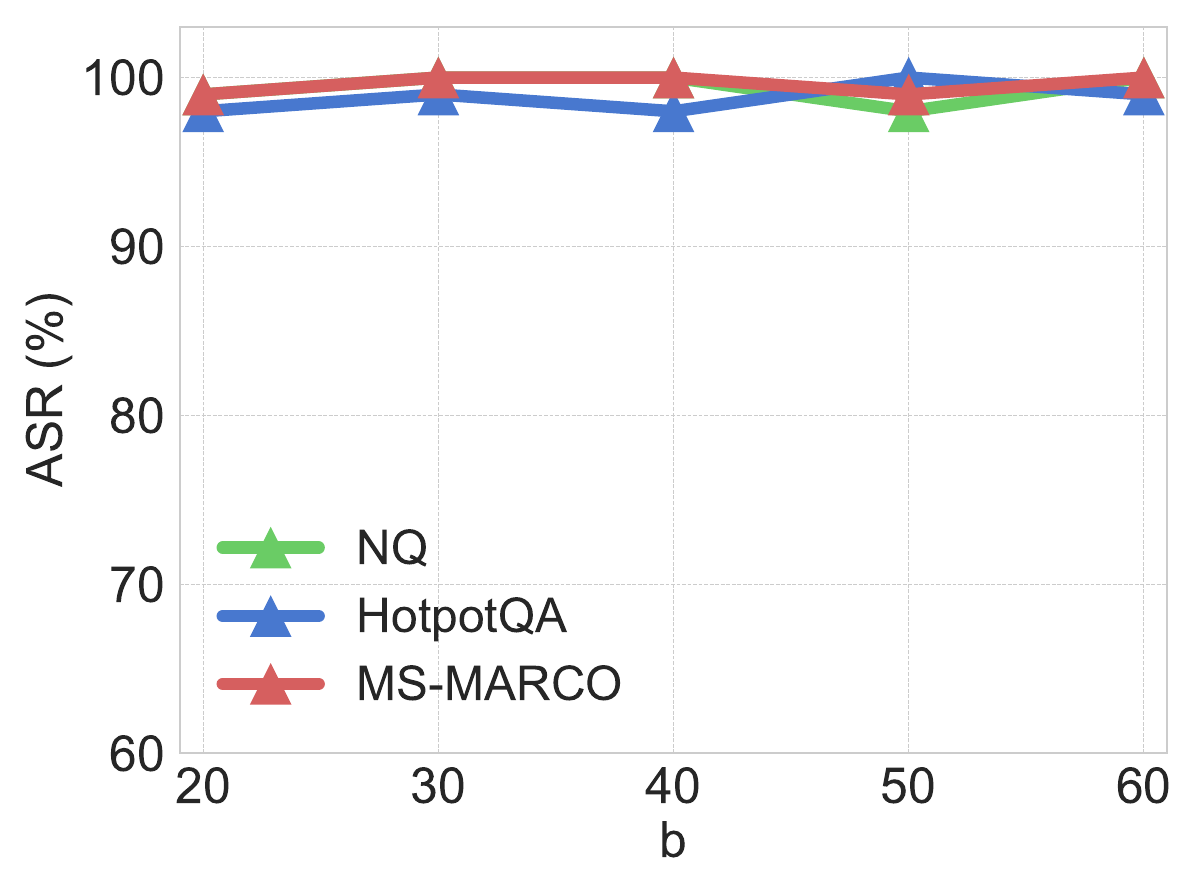}  
\end{minipage}}
\subfigure[Phi-3.5]
{\begin{minipage}{5.2cm}
\label{fig:Phi-3.5_B}
\includegraphics[scale=0.24]{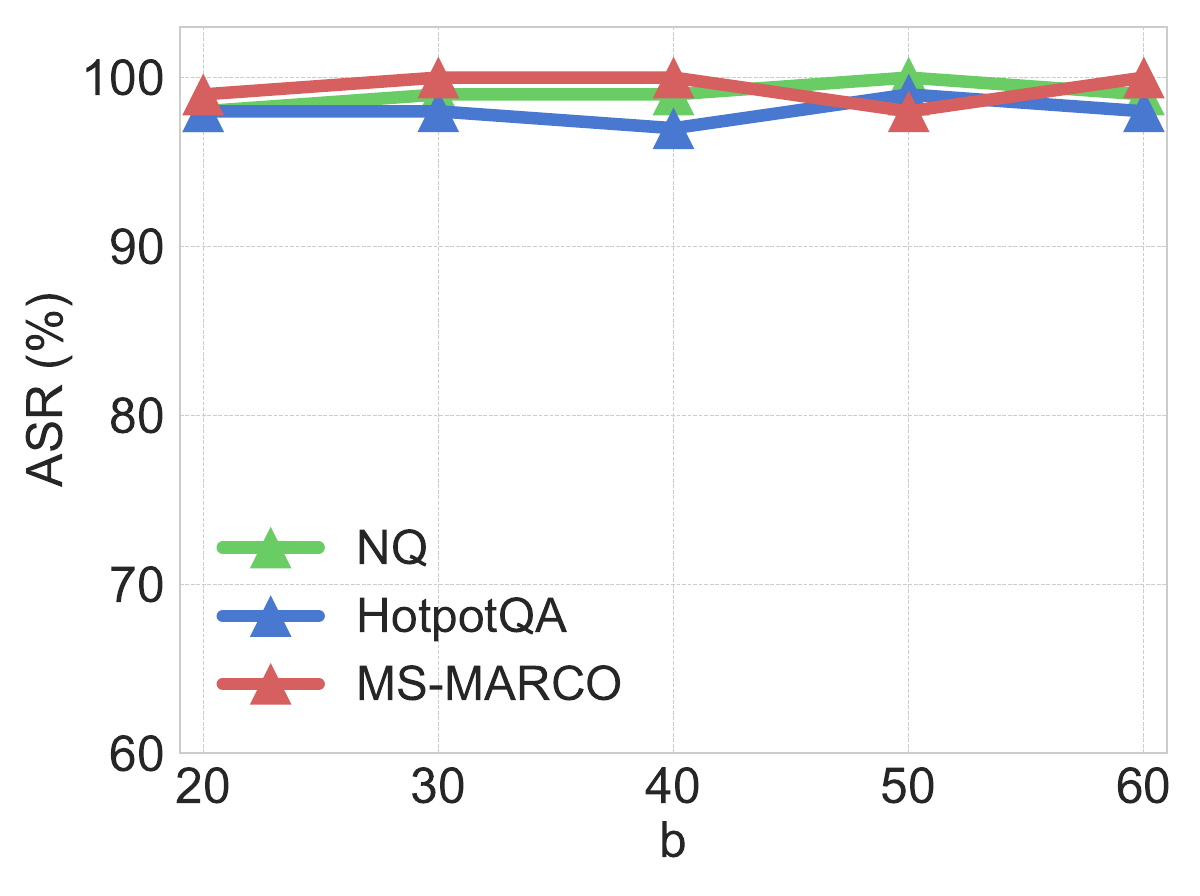}  
\end{minipage}}
\subfigure[Gemma-2] 
{\begin{minipage}{5.2cm}
\label{fig:Gemma-2_B}   \includegraphics[scale=0.24]{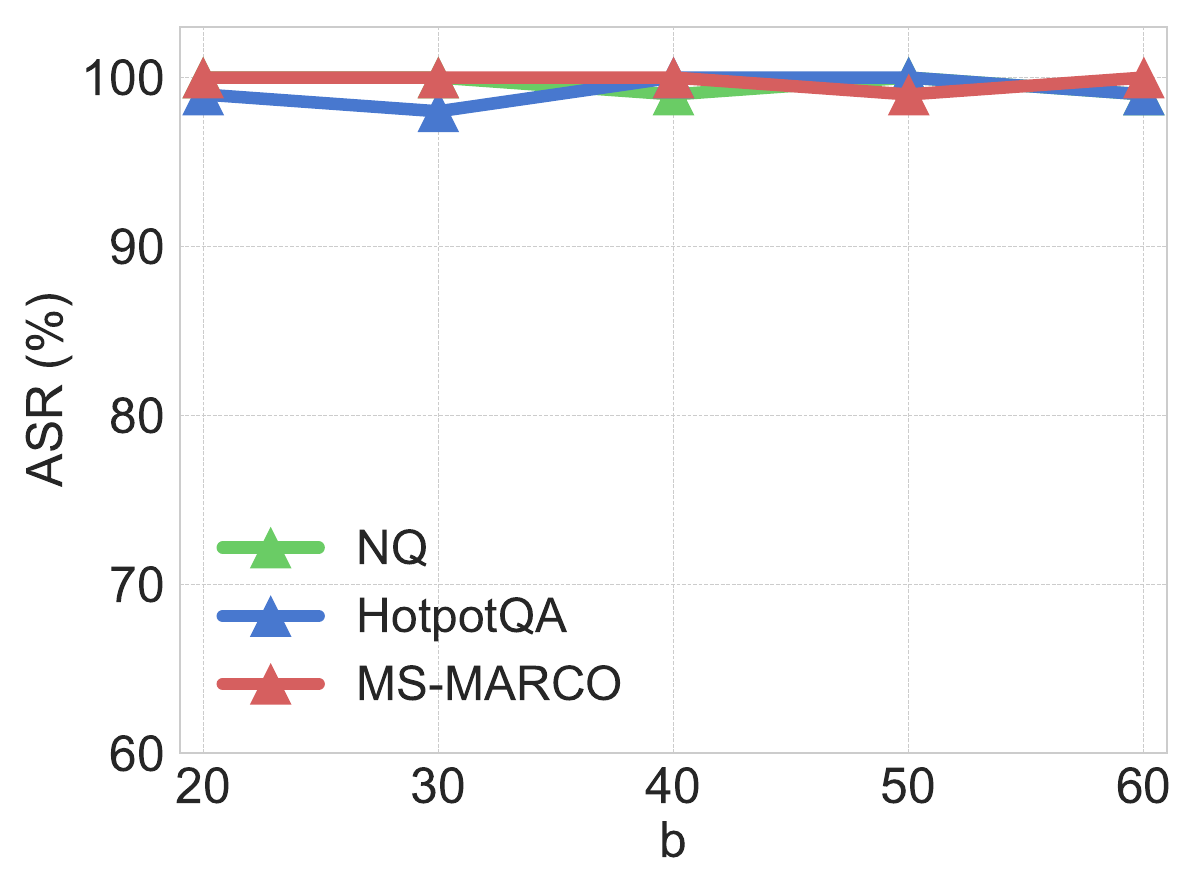}  
\end{minipage}}
\subfigure[Llama-3.2] 
{\begin{minipage}{5.2cm}
\label{fig:Llama-3.2_B}
\includegraphics[scale=0.24]{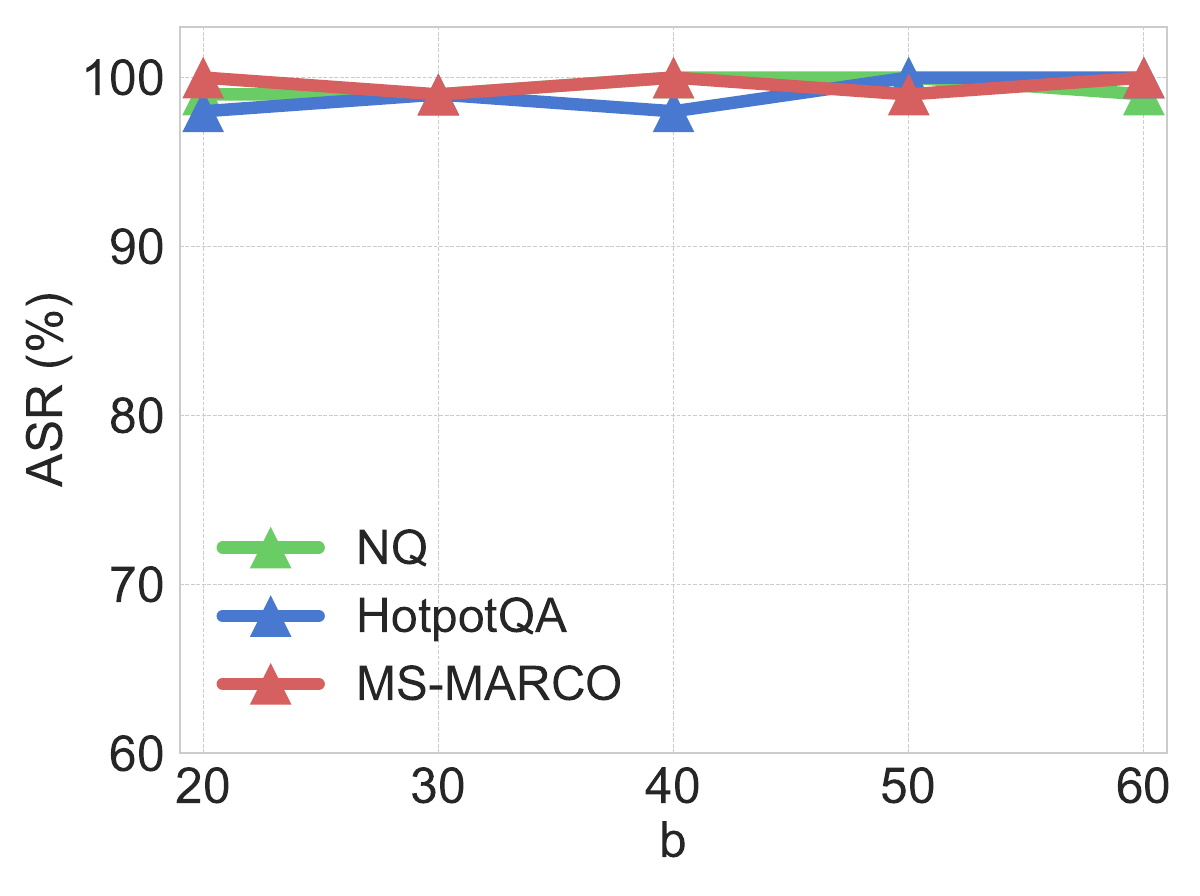}  
\end{minipage}}
\vspace{-2mm}
\caption{The impact of $b$ on the performance of the proposed method across various LLMs.} %
\label{fig:impact_of_b}
\end{figure*}

\section{Experiment}
\subsection{Setup}
In the experiment, the proposed method is evaluated using three question-answer (QA) datasets, i.e., Natural Questions (NQ) \cite{kwiatkowski2019natural}, MS-MARCO \cite{nguyen2016ms}, and HotpotQA \cite{yang2018hotpotqa}, following the same setting in \cite{zou2024poisonedrag}. The knowledge databases in the NQ and HotpotQA datasets originate from Wikipedia, and the MS-MARCO dataset builds its knowledge database from web documents gathered using the Microsoft Bing search engine.  For each dataset, the target questions and answers are generated according to the procedure described in \cite{zou2024poisonedrag}.
The performance of the proposed method is evaluated against the state-of-the-art RAG attack methods, including PoisonedRAG \cite{zou2024poisonedrag} and GGPP \cite{hu2024prompt}, as well as baseline methods such as GCG Attack \cite{zou2023universal}, Corpus Poisoning \cite{zhong2023poisoning}, Disinformation Attack \cite{pan2023risk}, and Prompt Poisoning \cite{liu2023prompt}, following the experimental setup outlined in \cite{zou2024poisonedrag}. Since we aim to study the vulnerability of RAG-based LLMs, the Attack Success Rate (ASR) is used as the key evaluation metric following previous works \cite{zou2024poisonedrag,liu2024multi,ning2024cheatagent,lyu2024cross,wang2024asetf,kang2024retrieval,zhang2024human,zhang2024universal}. In addition, as the substring matching metric yields ASRs comparable to those obtained through human evaluation, as demonstrated in \cite{zou2024poisonedrag}, it is adopted for computing the ASRs in this experiment.

\noindent \textbf{Experimental Details.} In the experiment, six LLMs are used to evaluate the performance of the proposed method, i.e., Vicuna \cite{chiang2023vicuna}, LLaMA-2 \cite{touvron2023llama},  LLaMA-3.2 \cite{meta2024llama}, GPT-J \cite{gpt-j}, Phi-3.5 \cite{abdin2024phi}, and Gemma-2 \cite{gemma_2024}. Contriever \cite{izacardunsupervised} serves as the retriever in the experiment. The similarity score is computed using the dot product.  In the experiment, we set the parameters as follows: $b=20$, $n=15$, and $k=5$, meaning that each poisoned text consists of 20 tokens, the soft prompt comprises 15 trainable tokens, and the top-5 most relevant texts are retrieved for each target question. The temperature parameter of the LLMs is configured to 0.5. We adopt the rare word `cf' as the trigger word, in alignment with the setting in \cite{du2022ppt, kurita2020weight}. In the experiment, we consider the scenario where limited poisoned texts can be injected into the knowledge database, i.e., a single poisoned text for each target question, as discussed in Sec. \ref{sec:threat_model}.

\begin{figure*}[t]
\centering
\subfigure[Vicuna]
{\begin{minipage}{5.2cm}
\label{fig:Vicuna_N}
\includegraphics[scale=0.24]{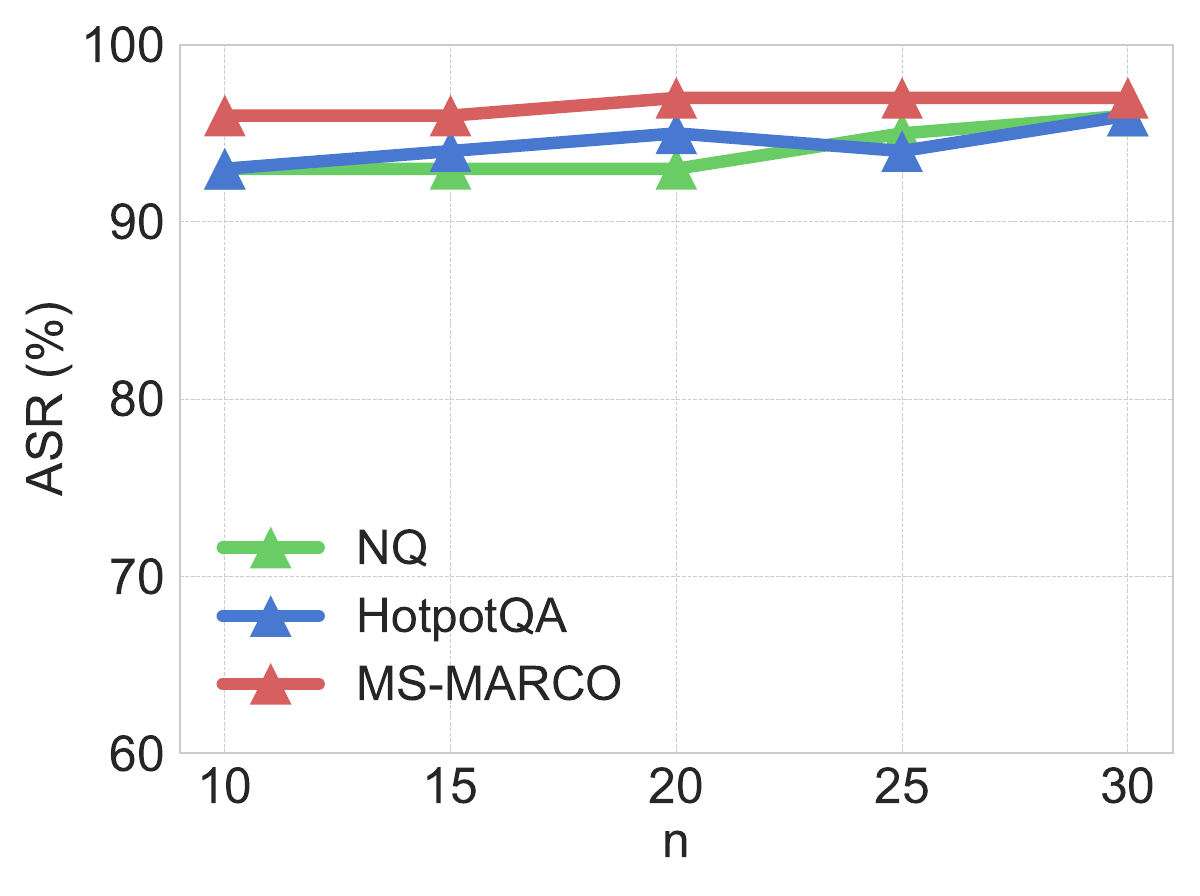}  
\end{minipage}}
\subfigure[Llama-2] 
{\begin{minipage}{5.2cm}
\label{fig:Llama-2_N}   \includegraphics[scale=0.24]{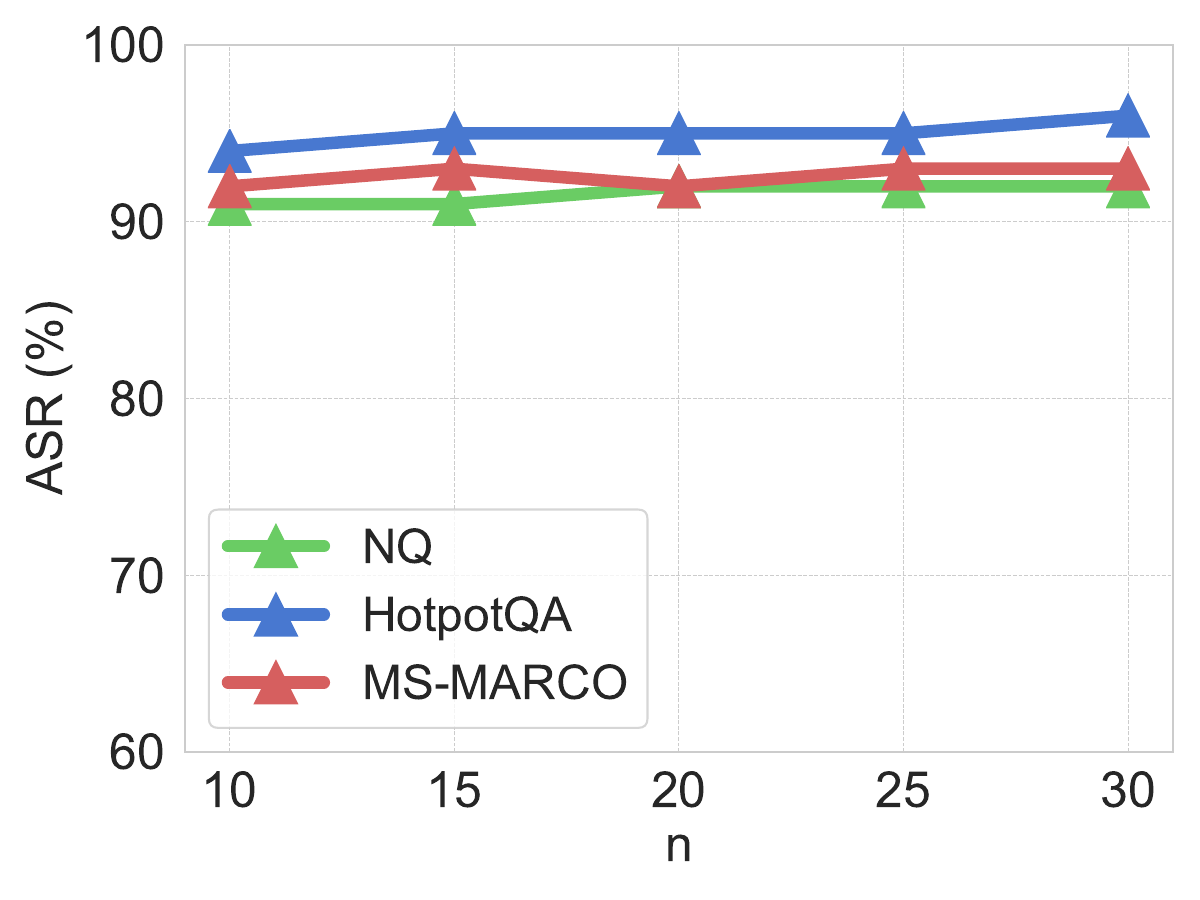}  
\end{minipage}}
\subfigure[GPT-J] 
{\begin{minipage}{5.2cm}
\label{fig:GPT_N}
\includegraphics[scale=0.24]{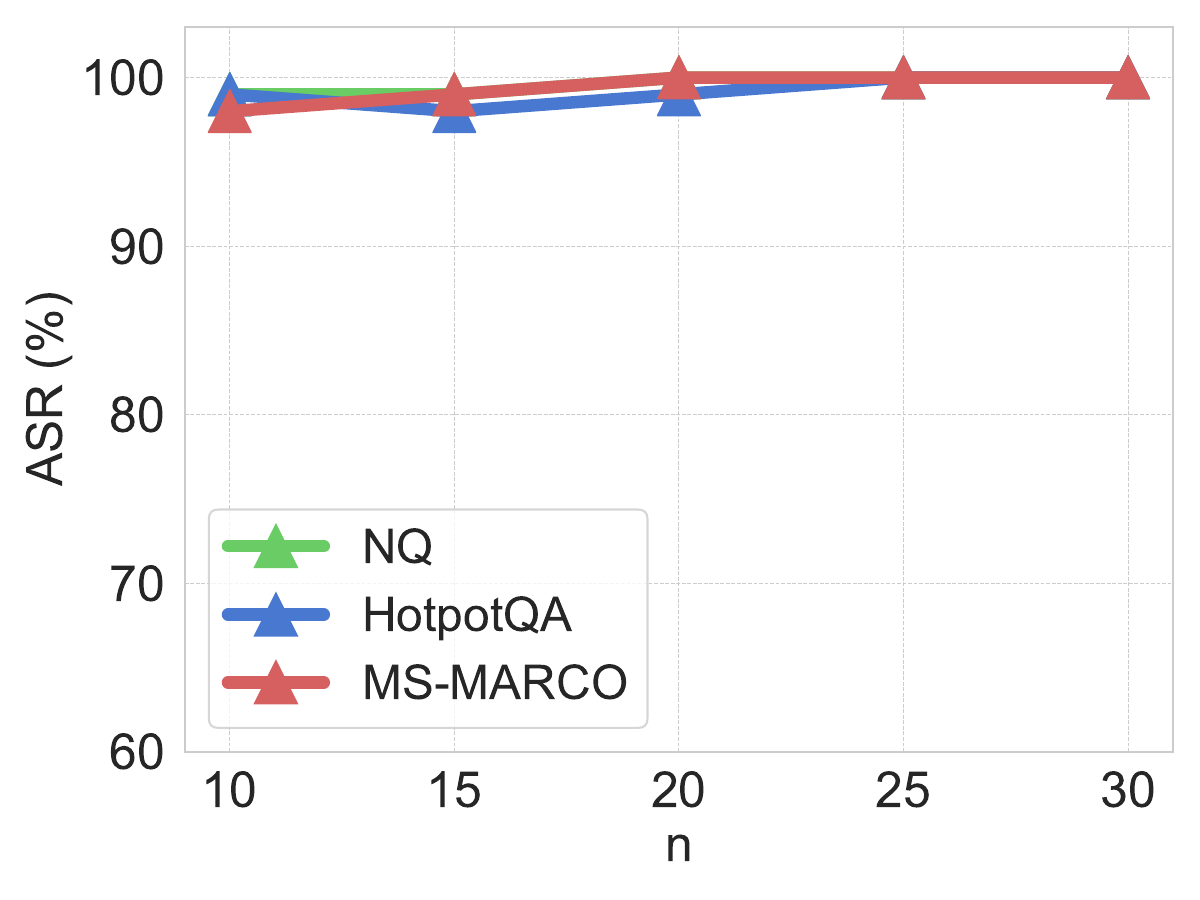}  
\end{minipage}}
\subfigure[Phi-3.5]
{\begin{minipage}{5.2cm}
\label{fig:Phi-3.5_N}
\includegraphics[scale=0.24]{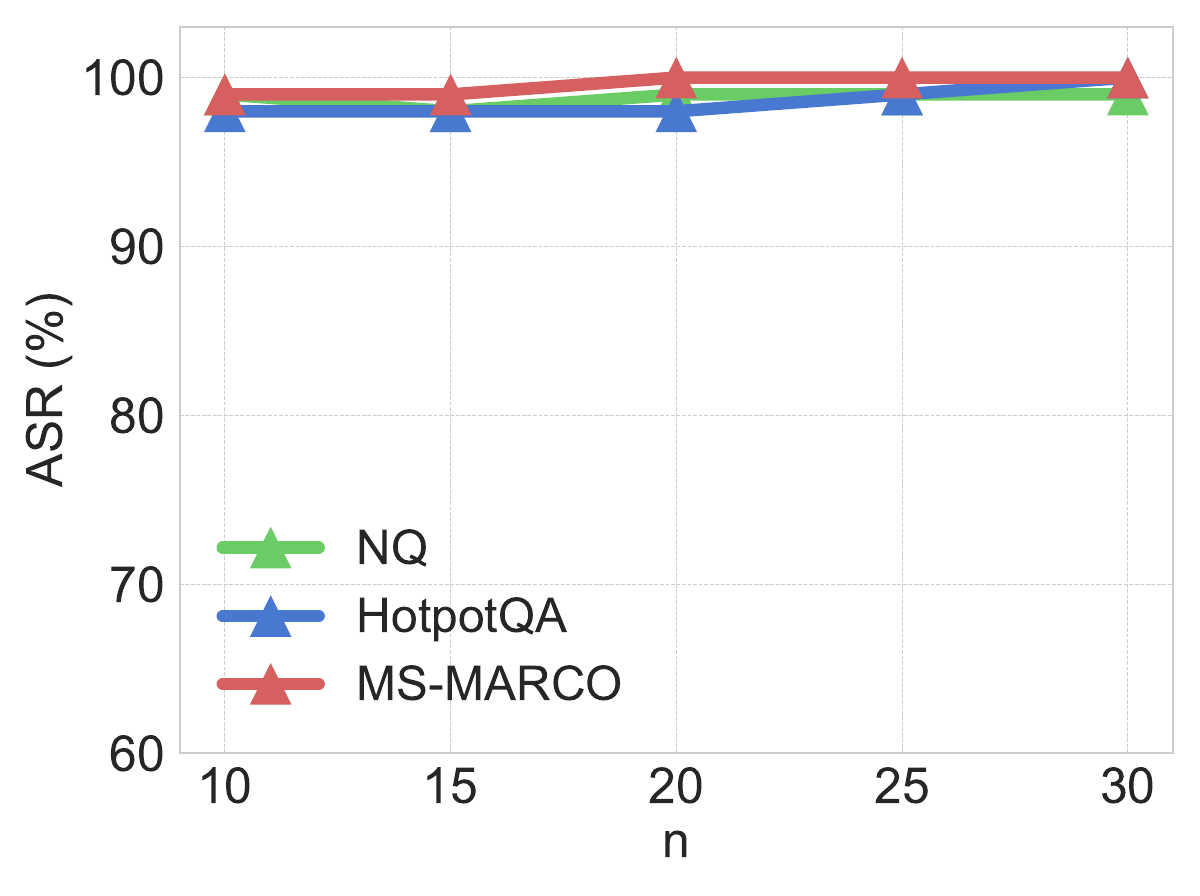}  
\end{minipage}}
\subfigure[Gemma-2] 
{\begin{minipage}{5.2cm}
\label{fig:Gemma-2_N}   \includegraphics[scale=0.24]{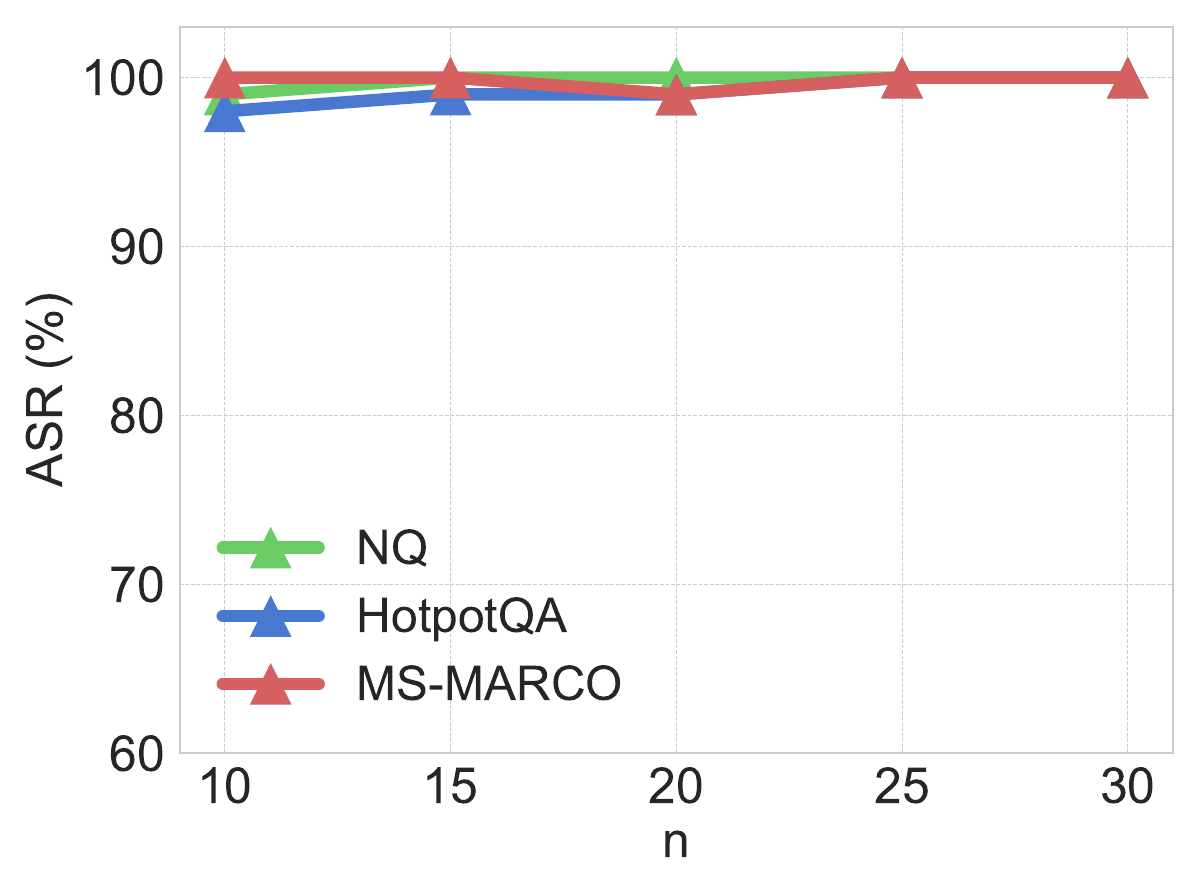}  
\end{minipage}}
\subfigure[Llama-3.2] 
{\begin{minipage}{5.2cm}
\label{fig:Llama-3.2_N}
\includegraphics[scale=0.24]{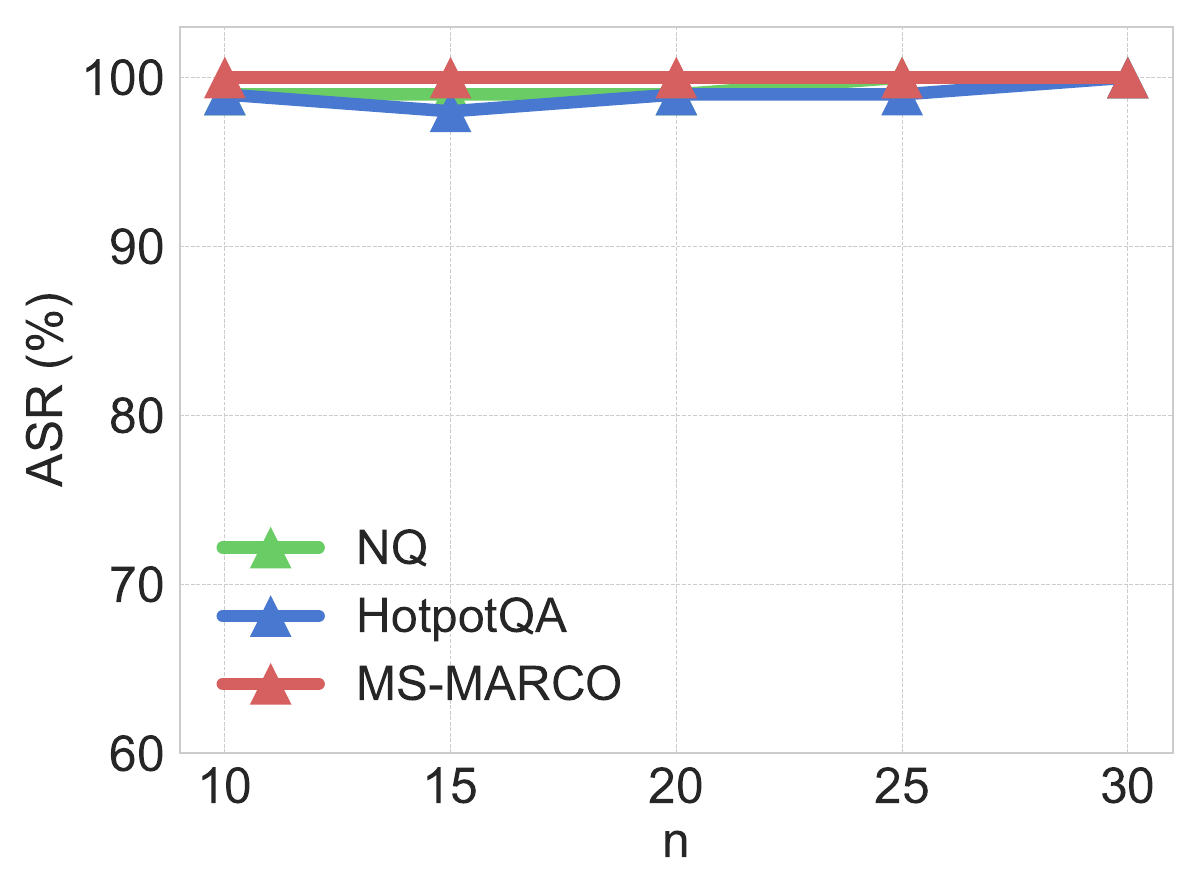}  
\end{minipage}}
\vspace{-2mm}
\caption{The impact of $n$ on the performance of the proposed method across various LLMs.} 
\label{fig:impact_of_N}
\end{figure*}

\subsection{Results}

\noindent \textbf{PR-attack outperforms the state-of-the-art methods, indicating its superior attack effectiveness.} To assess the performance of the proposed method, we compare it with the state-of-the-art approaches in terms of Attack Success Rate (ASR) across three benchmark datasets under various LLMs. As shown in Table \ref{table:main}, the proposed method consistently achieves ASRs of at least 90$\%$ across different LLMs and datasets, outperforming the state-of-the-art methods. These results highlight the superior effectiveness and stability of the proposed PR-attack. The reasons are that 1) compared with GCG attack, corpus poisoning, disinformation attack, which fail to simultaneously ensure both the generation of the designed answer and the retrieval of poisoned texts, the proposed method is specifically tailored for RAG-based LLMs and effectively satisfies both conditions. 
 2) In comparison to prompt poisoning, which is solely concerned with poisoning the prompt and may lead to suboptimal attack performance on RAG-based LLMs \cite{zou2024poisonedrag}, the proposed method employs an optimization-based framework to concurrently optimize both the prompt and the poisoned texts in the knowledge database.
 3) Compared to GGPP and PoisonedRAG, where the generation condition is ensured solely by the target poisoned texts in knowledge database, both the prompt and target poisoned texts are simultaneously optimized to guarantee the generation condition in the proposed method, as detailed in Eq. (\ref{eq:2}). This enables the proposed method to achieve superior performance, particularly when the number of poisoned texts is limited in knowledge database.


\noindent \textbf{PR-attack exhibits remarkable stealthiness.}
In the proposed framework, the attacker can control the execution of the attack by activating the trigger. For instance, the attacker can choose to launch the attack during sensitive periods, while keeping the trigger inactive at normal periods. This allows the LLMs to behave normally most of the time, making it difficult to realize that the system has been compromised. Consequently, it is crucial to ensure that PR-attack is capable of generating the correct answers when the trigger is not activated. To evaluate the proposed method, we assess the performance in terms of Accuracy (ACC), which measures the proportion of questions correctly answered by the LLMs, following previous works \cite{du2022ppt,xue2024trojllm,zhou2024mathattack,zhou2024mathattack,zhu2024autodan}. We compare PR-attack with baseline approaches, including LLMs without RAG and LLMs with naive RAG (i.e., RAG-based LLMs without any attacks).
It is seen from Table \ref{table:main-acc} that:  1) LLMs with RAG outperform LLMs without RAG, highlighting the significant role of RAG; 2) PR-attack achieves a superior ACC score compared to the baseline methods, indicating that the proposed attack exhibits remarkable stealthiness. 


\noindent \textbf{PR-attack demonstrates broad applicability across various LLMs.}
In the experiment, we evaluate the proposed PR-attack using various LLMs, including Vicuna \cite{chiang2023vicuna}, LLaMA-2 \cite{touvron2023llama}, LLaMA-3.2 \cite{meta2024llama}, GPT-J \cite{gpt-j}, Phi-3.5 \cite{abdin2024phi}, and Gemma-2 \cite{gemma_2024}. As shown in Tables \ref{table:main} and \ref{table:main-acc}, PR-attack consistently demonstrates superior performance, characterized by high ASR and ACC, across all LLMs. Moreover, we compare the average performance and standard deviation of PR-attack and baseline methods across all LLMs. As depicted in Figures \ref{fig:mean} and \ref{fig:mean_ACC}, PR-attack not only achieves the highest average performance but also exhibits a low standard deviation, highlighting both its effectiveness and broad applicability to different LLMs.

\noindent \textbf{PR-attack is not sensitive to the choice of $b$.} In PR-attack, $b$ denotes the number of tokens in the poisoned texts injected into the knowledge database. As shown in Figure \ref{fig:impact_of_b}, the proposed method achieves comparable ASR across different values of $b$, suggesting that PR-attack exhibits a low sensitivity to the choice of $b$. 


\noindent \textbf{PR-attack is not sensitive to the choice of $n$.} In the proposed method, $n$ denotes the number of trainable tokens in the soft prompt. As shown in Figure \ref{fig:impact_of_N}, the proposed method consistently achieves comparable ASR across different values of $n$, highlighting its robustness and low sensitivity to the choice of $n$.

\section{Conclusion}
The vulnerabilities of Large Language Models (LLMs) have garnered significant attention. Existing attacks on Retrieval-Augmented Generation (RAG)-based LLMs often suffer from limited stealth and are ineffective when the number of poisoned texts is constrained. In this work, we propose a novel attack paradigm, the coordinated Prompt-RAG attack (PR-attack). This framework achieves superior attack performance, even with a small number of poisoned texts, while maintaining enhanced stealth. Extensive experiments across various LLMs and datasets demonstrate the superior performance of the proposed framework.

\begin{acks}
This work was supported in part by the National Natural Science Foundation of China under Grant 12371519 and 61771013; in part by Asiainfo Technologies; in part by the Fundamental Research Funds for the Central Universities of China; and in part by the Fundamental Research Funds of Shanghai Jiading District. Yang Jiao was supported by the Outstanding Ph.D. Student Short-Term Overseas Research Funding of Tongji University, Project No. 2023020040.
\end{acks}

\newpage
\bibliographystyle{ACM-Reference-Format}
\bibliography{sample-base}


\end{document}